\documentclass[preprint,12pt]{elsarticle}

\usepackage{amssymb}
\usepackage{tabularx} 
\usepackage{url}

\journal{arXiv}

\begin{document}

\begin{frontmatter}



\title{The Security of Smart Buildings:\protect\\a Systematic Literature Review}

\author[lancsinfolab]{Pierre Ciholas\corref{pierre}\fnref{lancs}}
\ead{p.ciholas@lancaster.ac.uk}

\author[lancsinfolab]{Aidan Lennie\corref{aidan}\fnref{lancs}}
\ead{a.lennie@lancaster.ac.uk}

\author[kingslondon]{Parvin Sadigova\corref{parvin}\fnref{kings}}
\ead{parvin.sadigova@kcl.ac.uk}

\author[kingslondon]{Jose M. Such\corref{jose}\fnref{kings}}
\ead{jose.such@kcl.ac.uk}

\fntext[lancs]{Lancaster University, School of Computing and Communications}
\fntext[kings]{King's College London, Department of Informatics}



\address[lancsinfolab]{InfoLab21, School of computing and communications, Lancaster University, United Kingdom}
\address[kingslondon]{King’s College London, Department of Informatics, London, United Kingdom}

\begin{abstract}
Smart Buildings are networks of connected devices and software in charge of automatically managing and controlling several building functions such as HVAC, fire alarms, lighting, shading and more.
These systems evolved from mostly electronic and mechanical elements to complex systems relying on IT and wireless technologies and networks.
This exposes smart buildings to new risks and threats that need to be enumerated and addressed.
Research efforts have been done in several areas related to security in smart buildings but a clear overview of the research field is missing.
In this paper, we present the results of a systematic literature review that provides a thorough understanding of the state of the art in research on the security of smart buildings.
We found that the field of smart buildings security is growing significantly in complexity due to the many protocols introduced recently and that the research community is already studying.
We also found an important lack of empirical evaluations, though evaluations on testbeds and real systems seems to be growing.
Finally, we found an almost complete lack of consideration of non-technical aspects, such as social, organisational, and human factors, which are crucial in this type of systems, where ownership and liability is not always clear.
\\\hspace{\textwidth} 
\end{abstract}


\end{frontmatter}

\section{Introduction}

Smart Buildings (SB) are an interconnected set of sensors, actuators, controllers, devices and computers orchestrated together to provide and control the main functionalities of modern buildings.
SBs control subsystems such as Heating, Ventilation, Air Control (HVAC), water heating, lighting, shading, but also security-critical tasks such as CCTV, fire and intrusion alarms, and building physical access control \cite{bhatt2015design}\cite{sita2012building}.
SB were originally known as Building Management Systems (BMS), Building Control Systems (BCS), and Building Automation Systems (BAS) among others, but the increasing complexity of sensors and actuators, the variety of protocols and devices, the interconnection with other IT systems and the Internet, and the resulting advanced capabilities led to the so-called \emph{Smart} Buildings. SB are a type of Cyber Physical Systems (CPS), but with specific characteristics, protocols, standards, and technologies as we will see in this review, which deserves in-depth consideration and study.

Several reasons explain the widespread adoption of SB.
In addition to providing the aforementioned services, they also promise to increase energy savings, therefore reducing the carbon footprint. 
A good example are Building Energy Management Systems (BEMS) \cite{yoon2011distributed}, which use inputs from various subsystems such as occupancy detection and patterns, weather data in combination with indoor temperature, and humidity and lighting sensors to adjust or optimise heating, ventilation and air conditioning.
These systems individually reduce the overall energy consumed by the buildings, reducing the energy costs and the carbon footprint, but once connected together in a unified logic their efficiency can get significantly improved.
As an example, windows blinds can have a considerable impact on the HVAC control strategy \cite{kastner2005communication}. Also, SB aim at increasing comfort of inhabitants while reducing the maintenance costs. For instance, building managers can easily operate their buildings through user-friendly Human Machine Interfaces (HMI) from their computers and smartphones.
The remote control features of SB allow system integrators to reduce the cost of maintenance, by performing routine tasks remotely instead of sending an engineer on-site.
Furthermore, SB are also a strategic point in the Smart Cities paradigm \cite{gasco2018building}. For instance, 
integrating all buildings in a City Resources Management System (CRMS) could be used to balance the energy production and consumption, or reduce the number of power failures \cite{sita2014knx}. 

In order to offer all these functionalities mentioned above, 
SB increasingly use standardised and open technologies, often communicating through wired or wireless networks using several protocols. Some of these protocols are shared with other IT technologies and CPS, but others are specific of SB such as BACNet and KNX. 
Common wireless protocols used in SB include ZigBee, EnOcean, and Z-Wave among others. 
In addition, 
a large number of SB are directly connected and accessible through the Internet \cite{praus2014identifying} to provide integrators and users remote maintenance and control capabilities.
Besides technological details, commercial buildings usually host several companies or organisations.
This makes SB potentially much more directly connected to regular IT networks than other CPS such as Industrial Control Systems, where the engineers in charge of the manufacturing system network have a better overview of the interconnections and may have more options to segregate networks using specific cabling, firewalls, VLAN, air gaps, and other isolation solutions.
Even in cases where a building is in charge of one single company owning the whole facility, SB may fall out of the regular network security assessments and plans. 

Attackers are increasingly targeting SB as a potential way to compromise the security of companies and organisations, as they know SB can be a weak potential entry-point to break in or facilitate access to the network and/or the physical facilities of a company/organisation.
In 2013, a building belonging to Google that was accessible and vulnerable from the Internet was targeted, letting the attackers retrieve extremely detailed diagrams of the building along with access to multiple SBs features \cite{googlewharf7}.
The Miami TGK prison system also suffered of an incident that led to the system opening the cells doors of prisoners \cite{boyes2015best}.
BACNet devices have already been infected by botnet malwares and used to conduct distributed attacks \cite{vceleda2012flow}, and some researchers foresee smart building BotNets further arising in the future \cite{caviglione2015analysis}.

In this paper, we present the results of a systematic literature review on the current state of the art in research on the security of smart buildings. Based on this, we conducted both a general analysis to give a clear picture of the research conducted in the field so far categorising different types of related papers and providing quantitative measures of the evolution of the field over the years, as well as a detailed analysis of each of the approaches we found in the literature. Using both the general and the detailed analysis of the literature, we observed that: i) the field of smart buildings security is growing significantly in complexity due to the many protocols introduced recently and that the research community is already studying; ii) there is a lack of empirical evaluations, though evaluations on test-beds and real systems seems to be growing; and iii) there is an almost complete lack of consideration of non-technical aspects, such as social, organisational, and human factors, which are crucial in this type of systems, where ownership and liability is not always clear. Based on these findings, we propose a roadmap and a set of open challenges for future research on the security of SB.

\section{Background}


SB provide tools, hardware and software, designed to automate, monitor, and control mainly indoor but also outdoor building-related tasks such as \cite{kastner2005communication}:
1) Climate control, with HVAC systems including cooling/refrigeration, humidification, air quality control; 
2) Visual comfort, with artificial lighting, daylighting (motorised blinds/shutters); 
3) Safety, with fire alarm, gas alarm, water leak detection, emergency sound system, emergency lighting, CCTV; 
4) Security, with intrusion alarm, access control, CCTV, audio surveillance; 
5) Transportation, with elevators, escalators, conveyor belts; 
6) One-way audio, with public address/audio distribution and sound reinforcement systems; 
7) Supply and disposal, with power distribution, waste management, fresh water/domestic hot water, waste water systems; 
8) Communication and information exchange with data networking, PBX (private branch exchange)/intercom, shared WAN access; 
9) Sundry special domains, with clock systems, flextime systems, presentation equipments (e.g. video walls), medical gas, pneumatic structure support systems (for airhouses).
To carry out these tasks, SB use a large variety of sensors, actuators, controllers, and supervision software or devices working together.


Over the years SB  have bridged the gap between old and new technologies, from the original Building Management Systems, which allowed an automated control of electro-mechanic solutions using IT technologies \cite{so2001building}, to its progression towards the modern SB.
Originally, different types of buildings (commercial, residential, institutional, and industrial) used different proprietary programmable logic control solutions, which provided them with some security through isolation and obscurity. Nowadays SB increasingly use standardised and open technologies, often communicating through computer networks. In fact, official standard bodies exist for SB to organise their evolution and harmonisation \cite{kastner2005communication}, including: 1) ISO TC 205 Building Environment Design; 2) CEN TC 247 Building Automation, Controls, and Building Management; 3) CENELEC TC 205 Home and Building Electronic Systems, HBES; and 4) ISO/IEC JTC1 SC25 WGI Information Technology, Home Electronic System. Also, modern SB technologies are no longer confined to particular types of buildings but can be used in any type of large functional building. Note, however, that some technologies differ for very small networks, such as in Home Automation Networks (HAN), domotics, or modern Smart Homes, where the cost of building solutions would be too high for adoption.

\begin{figure*}[!ht]
\centering
\includegraphics[scale=.5]{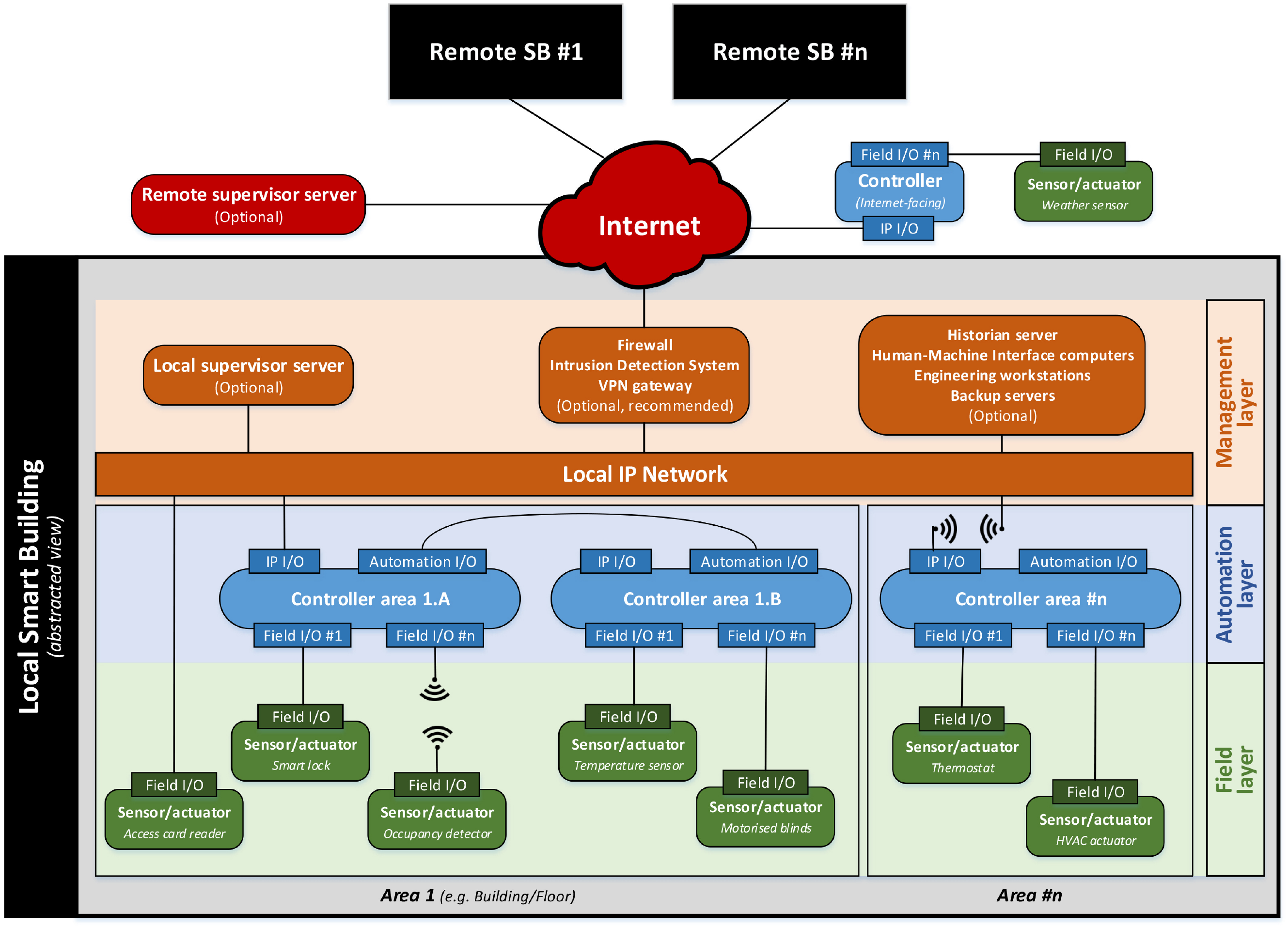}
 \caption{Abstracted layers of a SB \& Possible architectures of a large scale SB}
\label{fig:BMSdiagram}
\end{figure*}

An SB can be abstracted and divided in a set of 3 separate layers:
The field layer, the automation layer, and the management layer.
A diagram detailing these layers as well as the components possibly found at each of them and their possible interconnection is provided in Figure \ref{fig:BMSdiagram}.
It shows both a local-only SB (in grey) and a large scale possibly multi-building SB interconnected through the internet, with Internet-facing services and remote supervisor server.

The field layer is the closest to the physical building environment monitored and controlled.
Sensors and actuators are located at that layer and communicate with the controllers at the level above, the automation layer, using field layer specific protocols which can be wired or wireless. 
Some sensors and actuators communicate directly via simple electric signals --- e.g. a thermometer sensor sends a DC current of variable voltage indicating the temperature, the controller is then in charge of translating this tension into a human readable temperature information.
In some other cases, a sensor can directly communicate at the highest level, the management layer, over IP --- e.g. for building access control with access card scanners, the scanner is not connected to an intermediate controller but directly to the IP network communicating with the server managing the accesses.
Sensors and other devices at field level communicating directly with the management layer are becoming more and more common due to the increasing use of Internet of Things (IoT) in SB.
These devices are practically the fusion of both the sensor/actuator and the controller, they are therefore performing the required tasks of both the field and automation layer. This type of hybrid device is especially common in wireless IoT-type devices, examples include smart locks, smart electrical plugs, certain cameras or smoke detectors.

At the automation layer, different controllers, often denominated as Programmable Logic Controllers (PLC) or Direct Digital Controls (DDC) receive the information from nearby sensors.
They aggregate this data and can make it available to other controllers or servers over networks, or they can use this data using an internal logic to execute programmed tasks --- e.g. turning on/off an HVAC unit if the a room temperature sensor is above/below a threshold level.
Controllers at the automation level can communicate with other devices at automation level using protocols that are mostly SB-specific, such as BACNet or KNX for example, and can also use an IP network going through the higher management layer, or directly using an automation-layer-only protocol -- e.g. using the upcoming BACNet over ZigBee.
The devices at automation layer are the bridge between the cyber world and the physical world.
To make this bridge, they most of the time have inputs and outputs ports of various types to allow them to communicate with field level devices, other automation layer devices, and Ethernet or wireless ports allowing them to communicate over a classic computer network with one or several supervision servers. Note, however, that a supervisory server is not always required, controllers at automation layer most of the time function independently to handle their local tasks.
In fact, the overall SB control is traditionally not handled by the supervision servers, which makes these systems distributed, non-centralised, and to a certain extent resilient to network problems.

The data of all the controllers is centralised at management layer, where most of the time a local supervisor server, similar to a SCADA (Supervisory Control And Data Acquisition) server in other types of CPS, supervises the whole SB process, logs the activity, and more generally optimises the processes based on this data.
The controllers at automation layer can be connected over IP locally or directly to the Internet.
On the management level, IT standards prevail for the connectivity \cite{kastner2005communication}.
Therefore the SCADA server can be located off-site and can retrieve the data over the internet, but a local supervisor is the most general and safer case.
In addition to this supervisor, a set of other servers can be found at that layer, such as a historian that collects data over long periods of time often used in energy management systems, human machine interfaces to visualise the system state or modify it, and others.
The IP network used at management layer is often connected to the internet, possibly though security mechanisms such as firewalls, intrusion detection systems, or VPN gateway to filter access.
This connection to the internet enable remote management and maintenance, which allows to save time and money to the SB contractors and their customers.

Even though the 3 layer model presented above is useful to clarify the different SB layers from a functionality point of view and has therefore its pedagogical value, in practice, automation layer protocols function over IP many times, so that the line between automation and management levels is blurred. In fact, another abstracted model of SB exist and is very commonly used in the SB literature using only 2 layers: \emph{Backbone} and \emph{Field} layer. In this model, the automation and management levels are merged into the Backbone layer, and the controllers are the link between these 2 layers. Note, however, that following the 2 layer model there are several subtleties that are difficult to capture. For instance,   
a controller could be exploited locally or remotely over IP to gain its control with an attack originating from the management level, for example leveraging the absence of authentication in the protocol BACNet, but the access obtained would grant full control of the controller's functionalities at automation level only and may not be useful to further the attack at the management layer.
In the reminder of this review, 
%
we will use the 3 layers model for its clarity. However, since there are only few cases in which the distinction between automation and management layer matters, we will discuss these 2 layers together, pointing to the distinction when needed.


\section{Methodology}

A literature review aims at providing the reader with insights and give an overview of the state of the research in the field of interest as complete as possible.
When a literature review is conducted in an non-systematic way, it may give a distorted vision of the research efforts by discussing extensively the publications serving the purpose of the writer without mentioning other research approaches.
To provide a such an overview of the state of research in this particular field, we conducted a systematic literature review following a transparent and repeatable procedure as described in \cite{okoli2010guide}.

\subsection{Search procedure}

We conducted the gathering of the papers considering the key computer science and security publication databases: Web of Science, IEEE Xplore, ACM Digital Library, Scopus, and Usenix, during Q4 of 2016. 

To find publications relevant to our research, we created two sets of keywords.
The first set allowed us to narrow down the results to the field of interest, and it included: Smart Building(s), Building Management System(s), Building Control System(s), and Building Automation System(s). The second set allowed us to narrows down to the security aspect, and it included: Security, Threat(s), Vulnerability(ies), Risk(s), Privacy, Assurance, Access control(s).
Using the APIs offered by these databases, we found all the papers having at least one keyword of each set of keywords. Throughout this process, we gathered a total of \textbf{1697 papers} that we kept for the vetting phase.


\subsection{Vetting procedure}

For each paper, we extracted the title, authors, date of publication and abstract in order to proceed to a vetting phase.
To select the relevant papers to include in our SLR, we have defined a set of questions to answer after reading each paper's title and abstract:

\begin{enumerate}
	\item Does the paper imply that SB are its primary domain focus? 
	\item Does the paper orient or provide an insight into vulnerabilities of SB or the technologies used in them? 
	\item Does the study directly address vulnerabilities of SB or the technologies used in them? 
\end{enumerate}

To be included, the papers had to comply with question 1 AND ( 2 OR 3 ).
In order to limit the subjectivity of this process, each paper's title and abstract was read and vetted by two researchers independently of each other.
A total of 45 papers were considered by both researchers as relevant to include in the study.
After this, we measured the agreement between the two independent researchers using the prevalence-adjusted bias-adjusted kappa \cite{cunningham2009more}, which is an improved version of Cohen's kappa statistic for datasets with high prevalence index like ours --- i.e.,
there were many more papers that both researchers agreed were not relevant than the papers that both researchers agreed were relevant.
The result was 0.81, which means there was very high agreement according to \cite{landis1977measurement}.
Despite the high agreement, due to the huge amount of papers retrieved during the search process (recall there were 1697 papers), there were 160 papers remaining where the two researchers did not completely agree, i.e., one of the researchers said the paper was relevant and the other said it was not relevant or vice-versa.
For these cases, a third researcher was given the list of the 160 papers in conflict with their title and abstract.
Using the same instructions, and without knowing what the others researchers had decided about the papers, the third researcher gave a third assessment of relevance to these papers, which in the end meant that a further 45 papers were added as relevant to make a total of \textbf{90 papers} to review.

\subsection{Reviewing papers}

We then proceeded to the full retrieval of the 90 papers.
For each paper included in our study, we read and reviewed them using a review template allowing us to extract useful information for further analysis.
We extracted the titles, authors, article length, date of publication and access, the place of publication, the keywords that made us find them, our summary of the paper, an extraction of the most significant quotes, a critical review, the keywords associated with the paper, and a set of questions.
The questions for each paper were:

\begin{enumerate}
\item What does this paper describe?

Attacks / Vulnerabilities / Defences / Other (specify)
\item What does this paper propose?

Mathematical or formal model or framework / Security protocol / System or software / Experimental study/ Other (specify)
\item How is the proposal in this paper evaluated?

On a real system / On a testbed / Using simulation / Through mathematical proofs / No evalutation / Other (specify)
\item Which protocols does this paper discuss?

BACNet / ZigBee / LonWorks (LonTalk) / EIB/KNX / Other (specify)
\item Does this paper describe threats, vulnerabilities or risks that are: (1) Specific to SB only;  (2) Specific to CPS in general (including ICS, SCADA, embedded devices and others); (3) Shared with classic IT environments?
\end{enumerate}

\section{General Analysis}
\label{sec:analysis}
\begin{figure}[!ht]
\centering
\includegraphics[scale=.7]{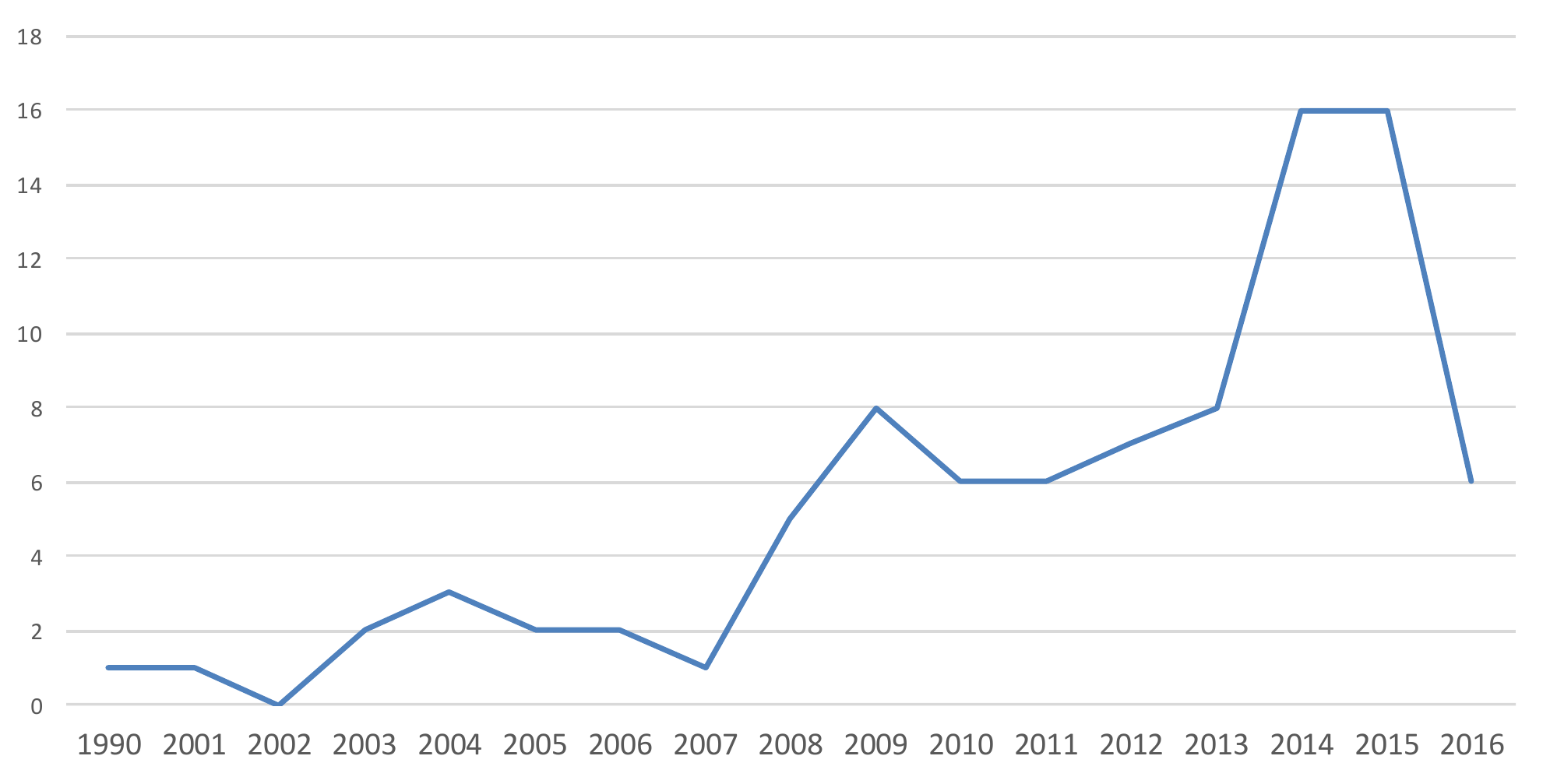}
\caption{Number of vetted publications over time}
\hspace*{-1.5in}
\label{fig:linegraph}
\end{figure}

\begin{figure}[!ht]
\centering
\includegraphics[scale=.5]{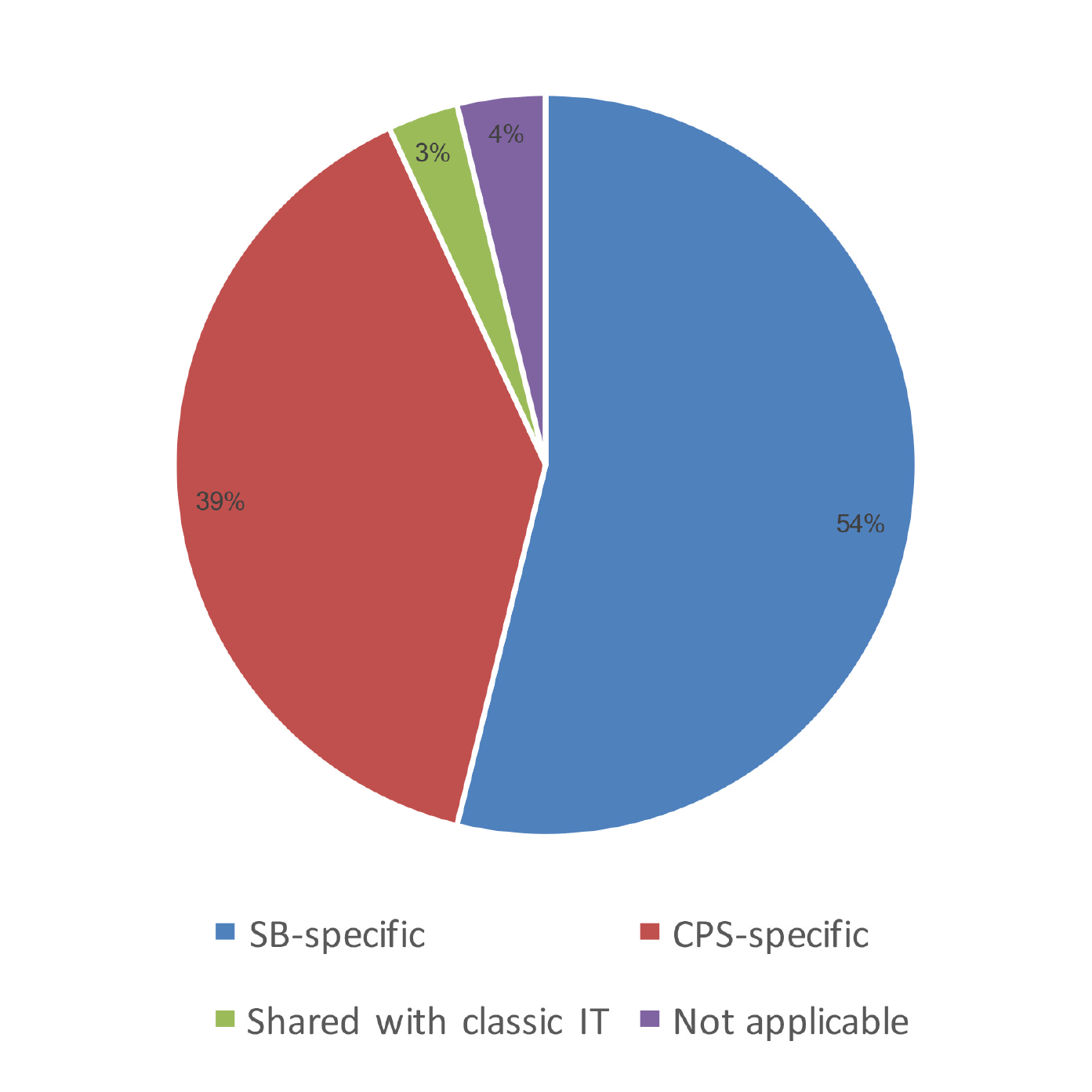}
\caption{Main domain considered per paper}
\hspace*{-1.5in}
\label{fig:fields}
\end{figure}

We first analyse the questions stated above to give a general overview of the field quantitatively.
Figure \ref{fig:linegraph} shows the number of vetted publications over time. Two clear periods can be observed, one going from 1990 to 2007 were the number of papers per year was very low, and after 2007 there is a steep increase over the years reaching a peak during 2014 and 2015.
Note that the period of collection was Q4 2016, which explains the difference in number between 2015 and 2016, as it takes some months for papers from acceptance to publication, and many papers may have been published in 2016 after our search. We can also see in Figure \ref{fig:fields} that most papers talk just about SB security, while some others also consider general aspects of cyber-physical systems in addition to particular SB issues, and very few of them address issues shared with regular IT systems. This highlights the importance of our review, as many of the issues discussed in the literature about SB security are indeed specific of SB and not of other systems.

\subsection{Main Topic}

\begin{figure}[!ht]
\centering
\includegraphics[scale=.5]{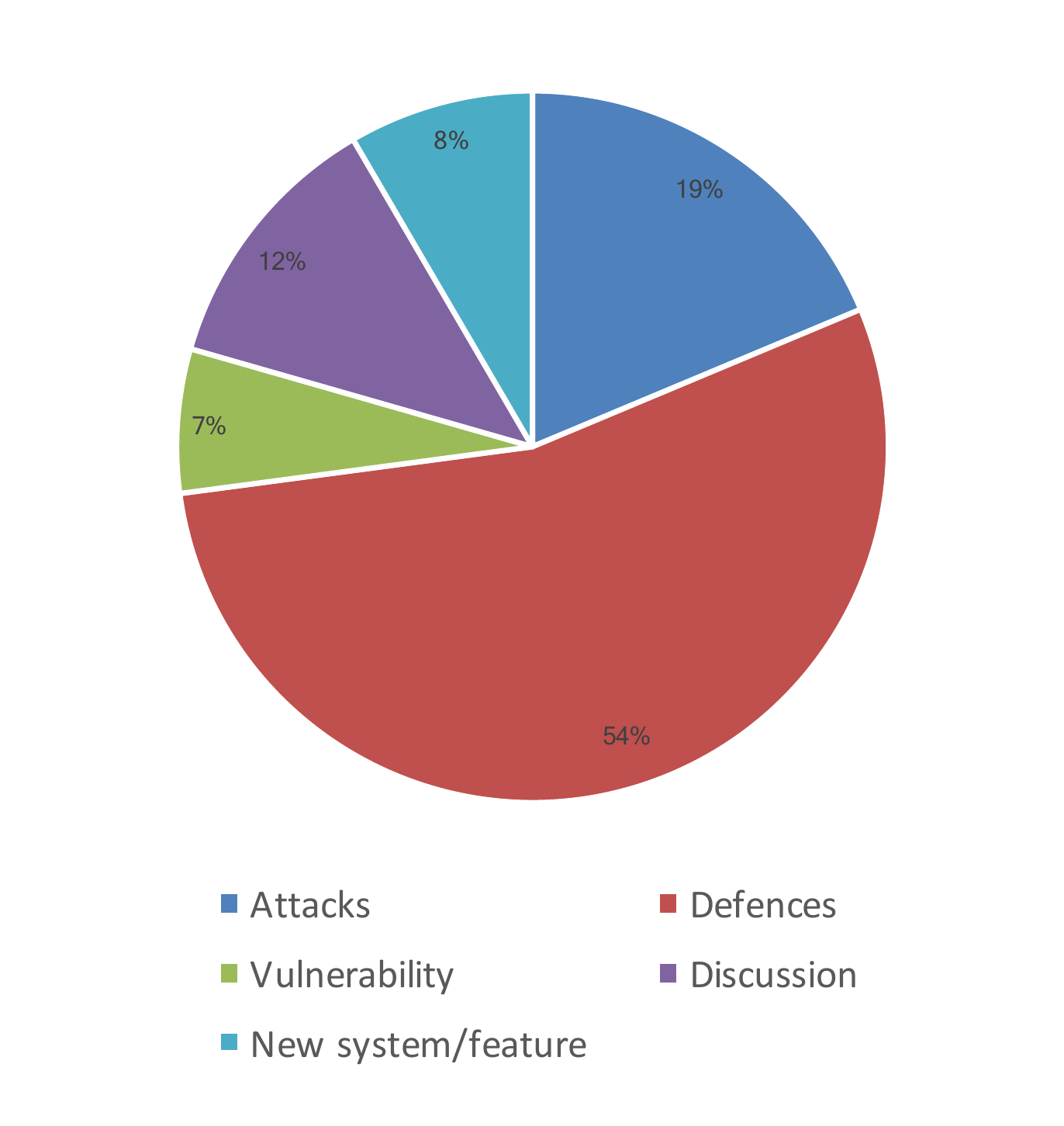}
\caption{Main Topic}
\hspace*{-1.5in}
\label{fig:presenting}
\end{figure}

As it is apparent from Figure \ref{fig:presenting}, the majority of the papers present defences (54\%).
The remaining half is divided between attacks (19\%), vulnerabilities (7\%), discussions (12\%) and a new systems/features(8\%).
However, note there were some overlaps of papers presenting both attacks and defences, which are accounted individually for each type.

\begin{figure}[!ht]
\centering
\includegraphics[scale=.62]{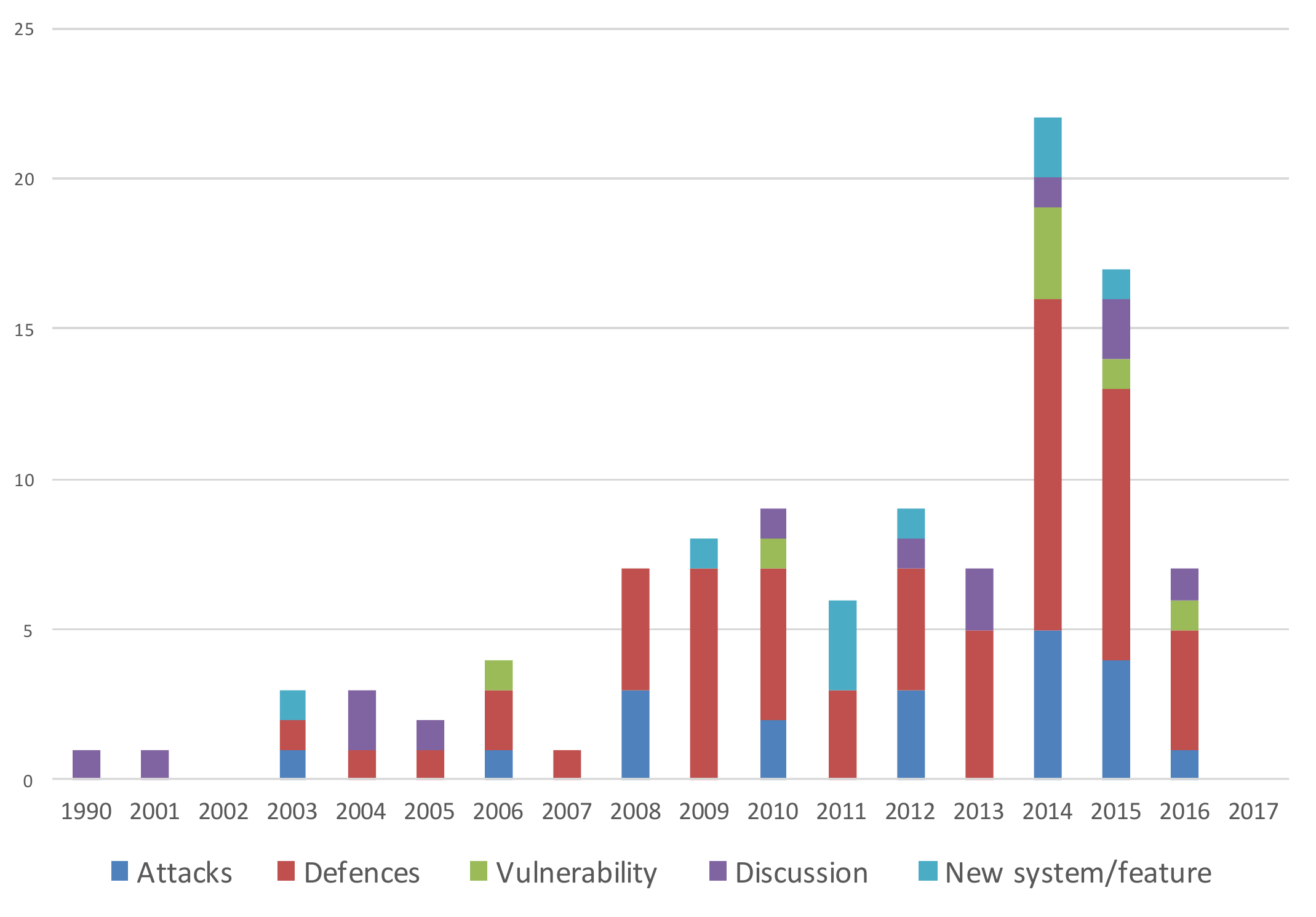}
\caption{Main topic over time}
\hspace*{-1.5in}
\label{fig:presentingbar}
\end{figure}

Figure \ref{fig:presentingbar} plots the main topic of the papers reviewed over time.
It is apparent that discussions about the topic were dominant in the early days, as initial explorations on the topic.
After that, the picture remains over time very similar to what can be observed in total in Figure \ref{fig:presenting}.
The split in the type of papers per year seems quite consistent with it.
The only exception seems to be that papers describing new systems/features seem to concentrate more in recent years, which makes sense considering that they could incorporate defences already developed in previous years.

\subsection{Protocols}
\begin{figure}[!ht]
\centering
\includegraphics[scale=.45]{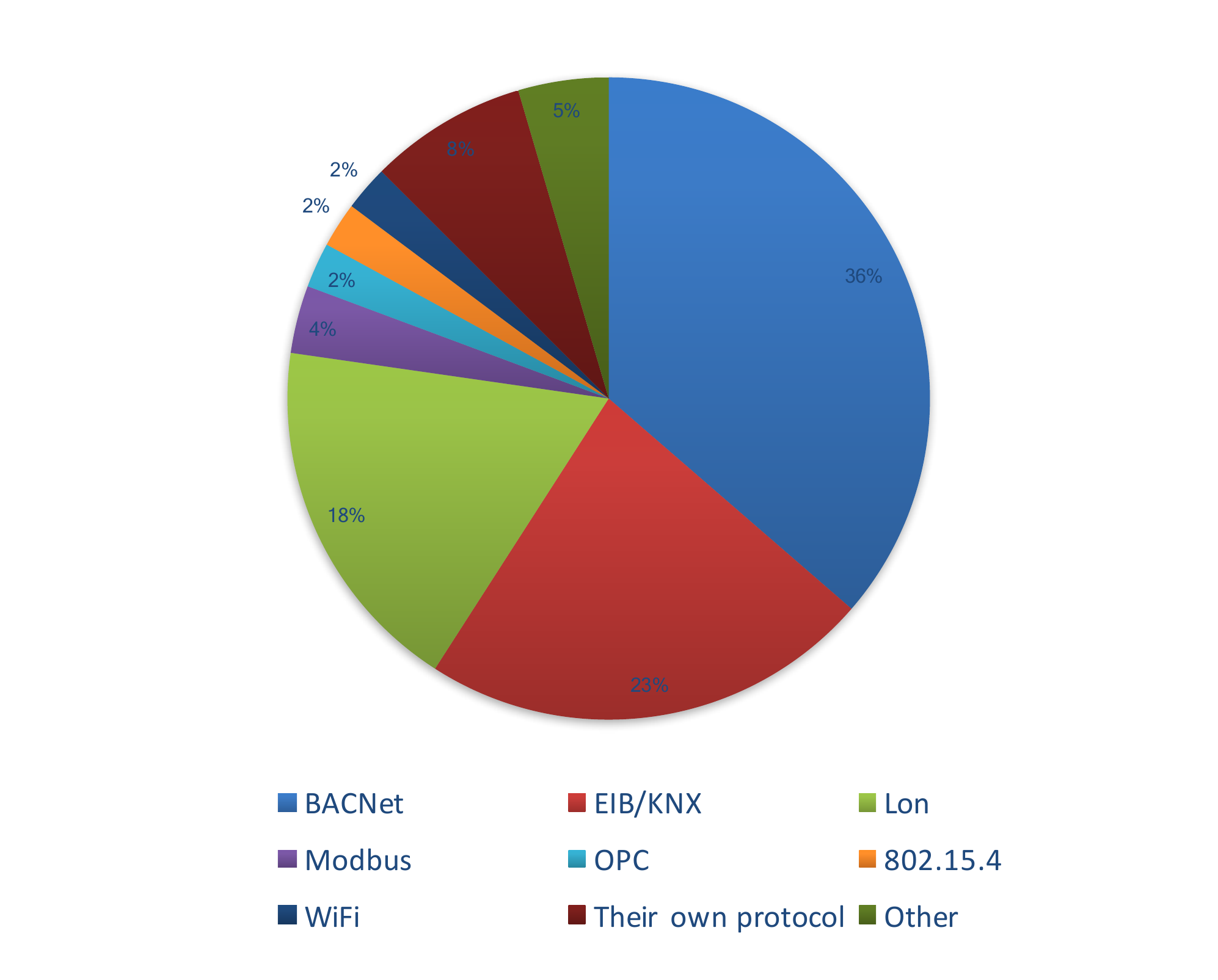}
\caption{Percentage of papers discussing protocols}
\hspace*{-1.5in}
\label{fig:protocols}
\end{figure}

The most popular SB protocols in the literature were clearly BACnet, KNX and Lon.
Figure \ref{fig:protocols} shows that 36\%, 23\% and 18\% of the papers discusses them, respectively.
Other protocols have been hardly considered, including 802.15.4 (the basis of Zigbee), Wi-Fi and OPC with around 2\% each.
Some of the reviewed papers discussed or proposed their own protocols (8\%).
Additionally, 5\% of papers mentioned other protocols including Z-Wave, WinCC, Step7 and others.

\begin{figure}[!ht]
\centering
\includegraphics[scale=.6]{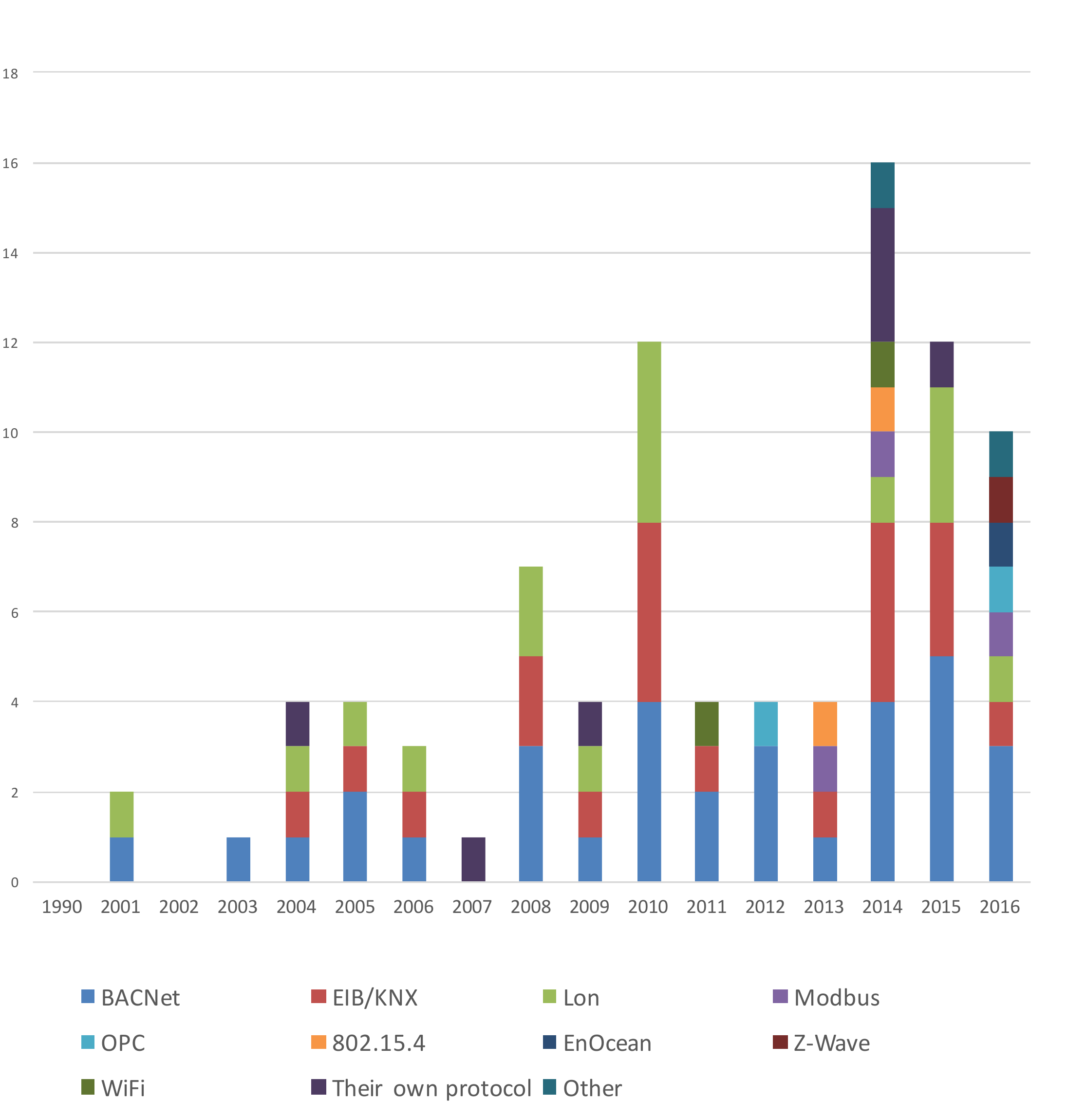}
\caption{Number of papers discussing protocols over time}
\hspace*{-1.5in}
\label{fig:protocolsbar}
\end{figure}

Figure \ref{fig:protocolsbar} illustrates the number of papers discussing particular protocols per year.
The dominance of BACnet, KNX and Lon can also be observed over time.
However, we can clearly see that Wifi-based protocols started to be discussed in 2011, and that the most recent protocols of this type, including 802.15.4 (the basis of Zigbee), EnOcean and Z-wave made it into papers from the last couple of years.
As a result, there is a very fragmented picture especially in 2014 and 2016, with still BACnet dominating, but with a clear increasing trend on the complexity of SB and the variety of protocols considered.

\subsection{Methodology}

\begin{figure}[!ht]
\centering
\includegraphics[scale=.44]{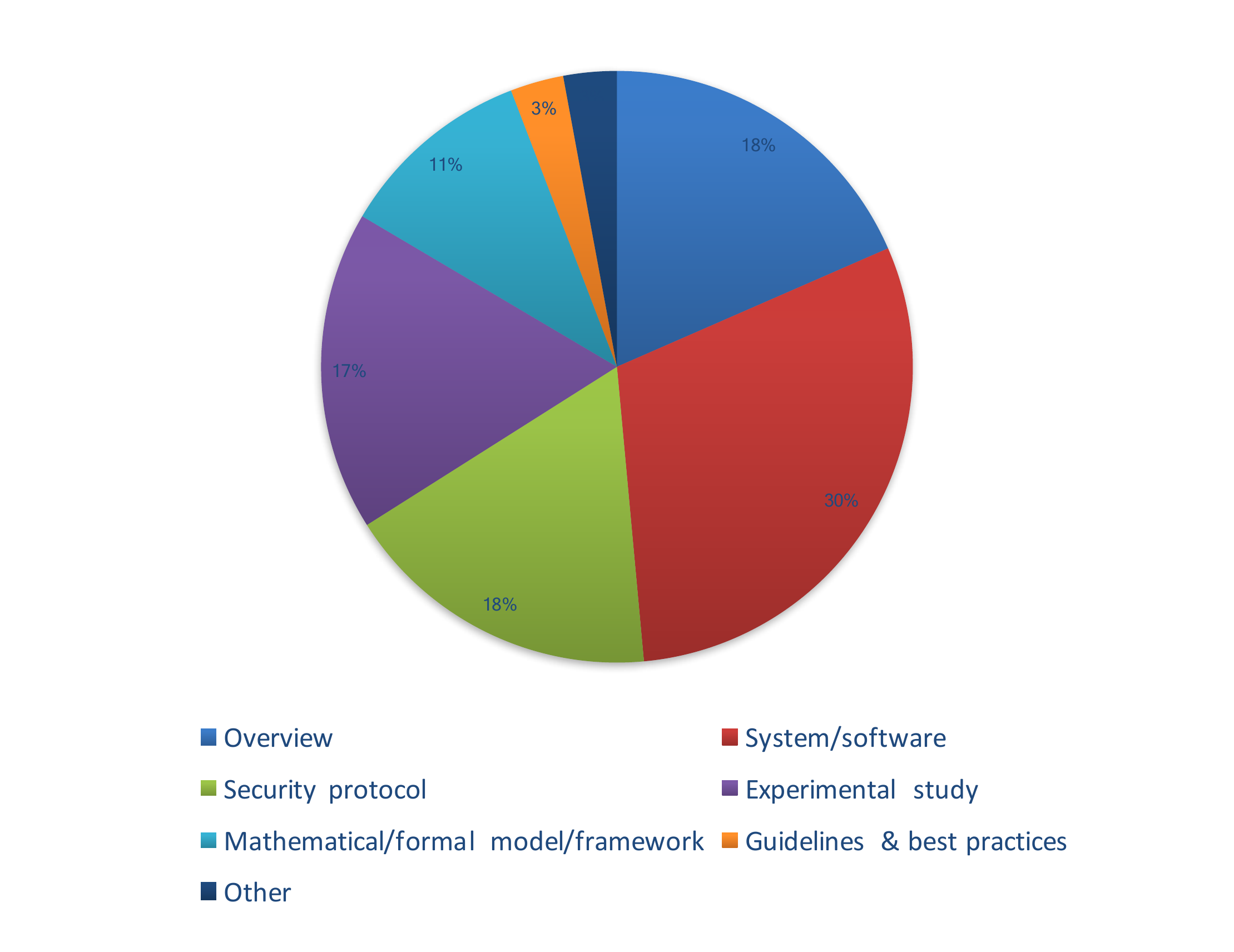}
\caption{Percentage of paper by methodology}
\hspace*{-1.5in}
\label{fig:proposing}
\end{figure}

As it is apparent from Figure \ref{fig:proposing}, the majority of the papers describe the design and/or implementation of a particular system/software (30\%).
Other methodologies include the design and/or implementation of security protocols (18\%), conducting experimental studies (17\%), and technology/literature overviews (18\%).
Formal papers working on / proposing new frameworks/ mathematical models constitute 11\% of the overall result.
Finally, a minority of papers describe guidelines/best practices (3\%).

\begin{figure}[!ht]
\centering
\includegraphics[scale=.58]{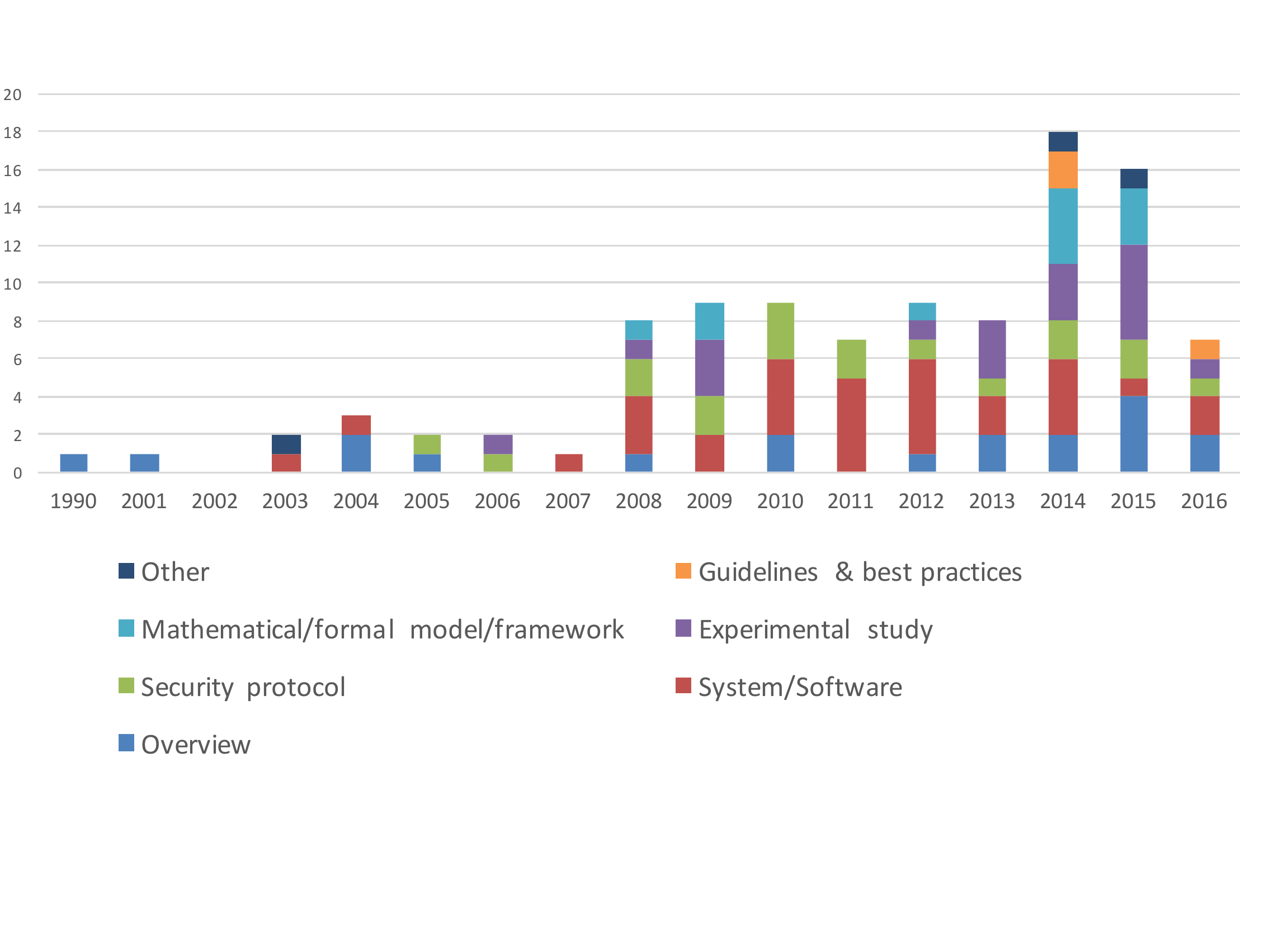}
\caption{Number of papers by methodology over time}
\hspace*{-1.5in}
\label{fig:proposingbar}
\end{figure}

Figure \ref{fig:proposingbar} illustrates the trend of papers with particular methodologies over time.
Generally speaking, the split of paper per year follows a very similar distribution to Figure \ref{fig:proposing}.
Notable exceptions are that overviews seem to dominate the early days, which would be consistent with the trend observed above about more high level discussions about the security issues of SBs, and the appearance of guidelines and best practices from 2014 on, which signals a maturing field of research that starts producing guidelines and best practices for practitioners.

\subsection{Evaluation}

\begin{figure}[!ht]
\centering
\includegraphics[scale=.55]{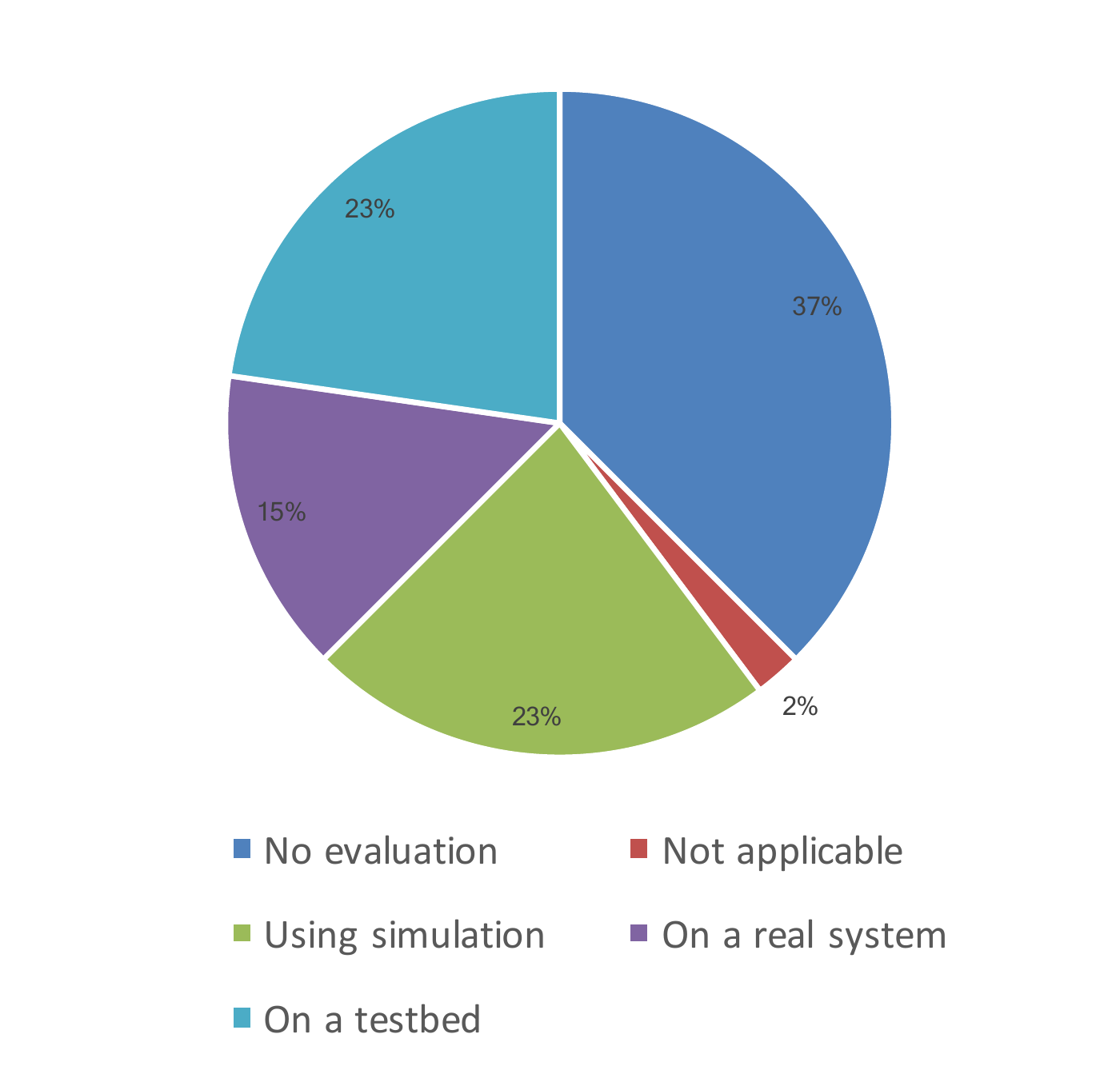}
\caption{Evaluation approach}
\hspace*{-1.5in}
\label{fig:Evaluated}
\end{figure}

\begin{figure}[!ht]
\centering
\includegraphics[scale=.64]{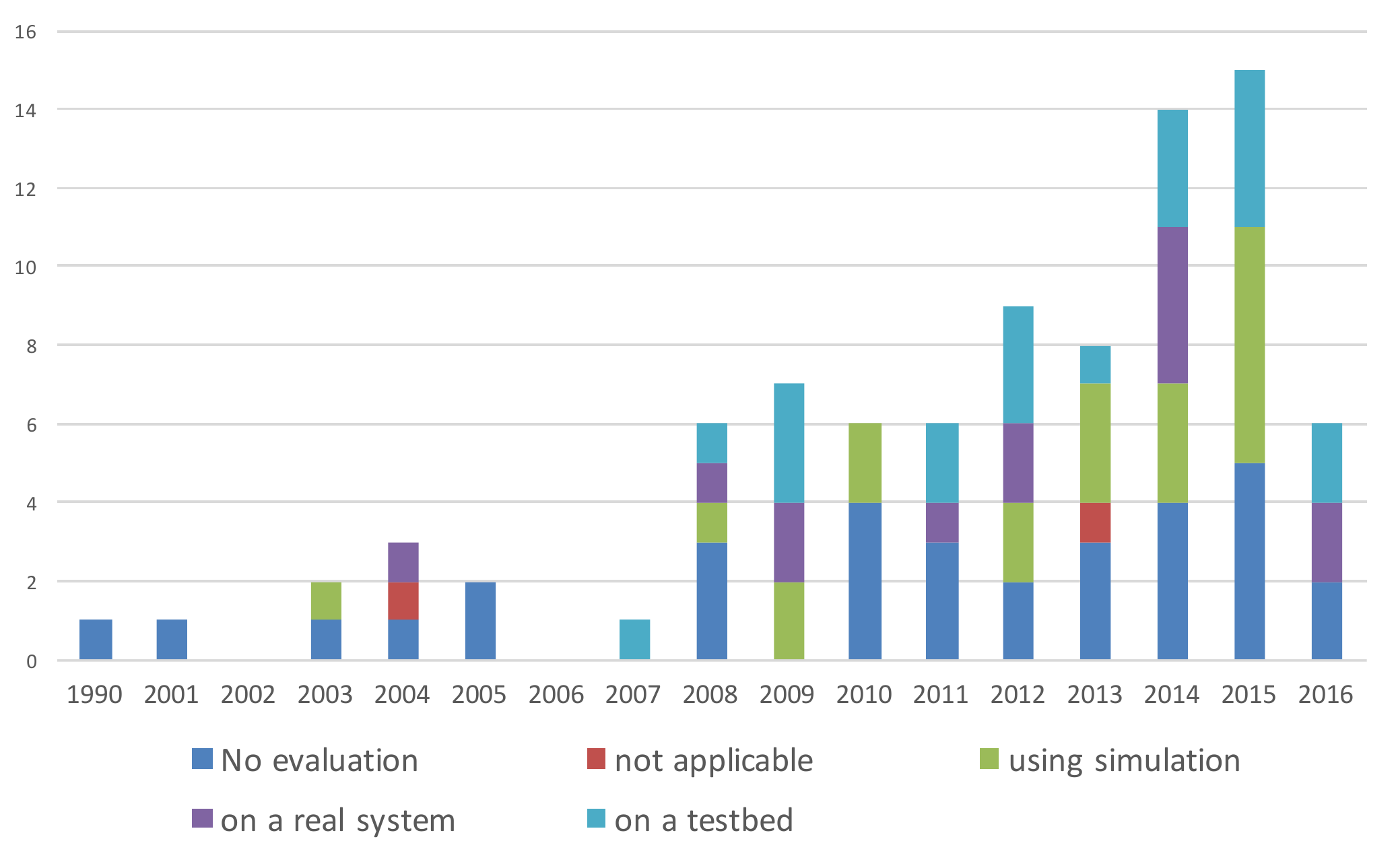}
\caption{Number of papers evaluated over time}
\hspace*{-1.5in}
\label{fig:evaluatedbar}
\end{figure}

Figure \ref{fig:Evaluated} shows the evaluation type followed by the papers reviewed.
The majority of them has no evaluation and that constitute 37\% from the overall result.
While some of this lack of evaluation could be attributed to some papers just being more of the type of discussions and guidelines, it may be a signal of the lack of a more empirical base for the analysis of security vulnerabilities and attacks and corresponding mitigations.
Especially worrying is the very low number of papers that have been evaluated on real systems (15\%), though conducting security studies of SB in the wild is obviously non-trivial. The picture of evaluation types over time in Figure \ref{fig:evaluatedbar} shows that evaluations have increased, particularly using real systems, testbeds, and simulations over the last 8 years. However, we can clearly see that a relatively high number of papers from the last couple of years still do not include any type of evaluation.

\section{Attacks and Vulnerabilities}

We now focus and give further details on the type of attacks and vulnerabilities discussed in the reviewed papers. We then report on defences in the following section. For both the attacks and defences, we will cover them by layer of abstracted SB in the following sections. Also note that, to facilitate the analysis, even if we separated discussion and new feature/system papers from attacks, vulnerabilities, and defences papers in the main topic of general analysis in the previous section to give a broad overview of types of papers, we now include them either in attacks and vulnerabilities and/or defences as appropriate, as many of them discuss potential vulnerabilities or incorporate defences as part of the new feature/system proposed --- even though that may not be the main topic of the paper. 

Table \ref{table:attacks} shows an overview of the papers that talk about SB attacks and vulnerabilities categorized into the SB layer they apply to and grouped around similar types of attacks and vulnerabilities. 
%
%
The publications focusing on attacks and vulnerabilities at field level are mostly focused on wireless systems (8/10 papers).
This is understandable considering that the security of wireless communications is an important field of cyber security research beyond SB. 
At automation and management layer, most attacks and vulnerabilities discussed fall in the category of denial of services (10 papers), followed by protocol-specific attacks (7 papers), configuration and legacy vulnerabilities (4), and privacy attacks (3).
Denial of services are one of the most common, as 
SB often have constrained computational power and are therefore more vulnerable toward this type of attacks.
Attacks targeting specific protocols are also numerous and can be explained by the evolution of these protocols, and more generally SB. The main protocols used in SB were not conceived with security in mind, since these networks were historically isolated.
Therefore, many protocols lack the most basic security mechanisms, and implementing them now that these protocols are deployed and used present numerous technical and financial problems.
Technically, implementing efficient security measures in the majority of the cases means having new devices incompatible with older versions of the protocol.
SB having life cycles spanning over decades and being managed by several different social entities and companies, upgrading to a secure solution is often either too complicated or too costly, in terms of both time and in money.
An excellent example is the BACNet protocol that was not developed with any sort of authentication mechanism.
Added to this lack of authentication, a feature named BBMD (BACNet Broadcast Management Device) allows to connect to a BACNet device via different buses including Ethernet/IP, and access the subdevices connected to the front device.
As a consequence of this, internet-facing BACNet devices without authentication and BBMD enabled may be easy to compromise and take control of. 


\begin{table}[!h]
\footnotesize
\begin{tabularx}{\textwidth}{|c|X|}
\hline
\multicolumn{2}{| c |}{\begin{large}\textbf{Attacks \& Vulnerabilities}\end{large}} \\
\hline
\multicolumn{2}{| c |}{\textbf{Field layer}} \\
\hline
Physical Attacks (2) &
\cite{hager2012secure}
\cite{mundt2016security} \\ 

\hline
Wireless Attacks (8) &
\cite{morgner2016all}
\cite{al2015anomaly}
\cite{granzer2010future}
\cite{praus2008secure}
\cite{cleveland2008cyber}
\cite{misra2010detection}
\cite{muraleedharan2006cross}\cite{sun2014review}
\\

\hline
\multicolumn{2}{| c |}{\textbf{Automation \& Management Layer}} \\

\hline
Denial of Service Attacks (10) &
\cite{granzer2008denial}
\cite{judmayer2014security}
\cite{kaur2015securing}
\cite{antonini2014practical}
\cite{caviglione2015analysis}
\cite{vceleda2012flow}
\cite{hager2012secure}
\cite{mundt2016security}
\cite{hill2004middleware}
\cite{cleveland2008cyber} \\

\hline
Protocol-specific Attacks (8) &
\cite{granzer2010future}
\cite{granzer2010security}
\cite{vceleda2012flow}
\cite{antonini2014security}
\cite{granzer2006security}
\cite{wendzel2012covert}
\cite{judmayer2014security}
\cite{praus2014identifying} \\

\hline
Configuration \& Legacy Vulns (4) &
\cite{hager2012secure}
\cite{boyes2015best}
\cite{zito2014decoding}
\cite{mundt2016security}\\

\hline
Privacy Attacks (3) &
\cite{mundt2016security}
\cite{caviglione2015analysis}
\cite{wang2014non}\\

\hline
Others (8) &
\cite{wei2015security}
\cite{fisk2012cyber}
\cite{mylrea2015cyber}
\cite{tankard2015secure} 
\cite{mansfield2015dangers}
\cite{hutt1990control}
\cite{lee2013reliable}
\cite{bo2014usable}\\

\hline
\end{tabularx}
\caption{Attacks \& Vulnerabilities categorized according to the SB layer and type of attack/vulnerability. \label{table:attacks}}
\end{table}

\subsection{Field layer}

We start with attacks on the field layer --- i.e., attacks on the sensors, actuators and controllers in SB that interact with the physical building environment and then connect to other sensors, actuators or controllers. The devices at field layer are the senses of the SB, the situational awareness of the entire SB relies on the good measurement of the physical information, therefore tampering with sensors can allow to modify the system's operational behaviour. 


\subsubsection{Physical Attacks} 

Very few papers consider the issue about physical attacks on sensors, actuators, and controllers in SB as they require physical access to them, which may not be possible in most of the cases and/or the access to these devices might be controlled.
For instance, in \cite{hager2012secure}, the authors identified potential physical attacks like vandalism, burglary, and theft of SB devices. 
A brilliant example of physical attacks at field layer is demonstrated in \cite{mundt2016security}.
The researchers present the potential physical damages of a SB attack such as mechanical destruction, electrical overload, premature ageing of the system, direct damage, locking doors and people, air conditioning and lighting control modification which may cause panic, among other effects.
The researchers could connect to the KNX network from a simple connected light switch, or from a USB plug left at several locations for maintenance purpose by the installation team. Interestingly, the researchers once connected to the network could read and write all types of telegrams from the entire system and hence fully control it. 
In their experiment, they were able to switch lights on and off, change temperatures, control windows blinds and HVAC.
The researchers could also read all the traffic in the KNX bus without even touching any cables through the walls.
As KNX over twisted pair cables uses 9600 baud, the frequency is within the range of audible area. 
Therefore, they used an audio pre-amplifier and a recorder.
They could then transform the traffic from their audio recording to decode all telegrams.
This allowed them after analysis to create a list of devices and their functionality being used on the network.
From this data they could also track people in the building and know how much time they spent in specific areas.
A very interesting observation is as KNX works on both the field layer and the automation layer, allowing packets to travel from controller to controller, the researchers could send messages on the automation layer from the field layer.

\subsubsection{Wireless Attacks}

Many of the sensors, actuators and controllers in SB now have wireless capabilities, which simplify installation and maintenance, but that actually introduce other potential attacks beyond physical ones, as the attacker does not need to have direct physical access to the devices but only be close enough to them.
We would like to point out that, interestingly, we did not encounter any publication presenting an attack allowing an attacker to start from an attack at the field layer then go up to the automation or management level.
Even in the case of the recent attack on the wireless ZigBee light bulbs by Philips \cite{morgner2016all} that allowed the attackers to take control of these devices, they could finally use their ZigBee radio transmitter to contaminate other bulbs or eventually jam the radio communications but they could not infiltrate or contaminate beyond the ZigBee network.
Because of the difficulties to use the field level as an attack vector for a higher objective, most wireless attacks described in the literature aim at disrupting or interrupting operations at this level, such as \cite{al2015anomaly}, \cite{muraleedharan2006cross}, \cite{granzer2010future}, \cite{sun2014review}, \cite{misra2010detection}, \cite{praus2008secure}, \cite{cleveland2008cyber}. 
The most common DoS attacks on ZigBee are presented in \cite{al2015anomaly}, including flooding, jamming, and pulse DoS. DoS attacks aimed specifically at sensor nodes are highlighted in \cite{muraleedharan2006cross}, where by launching wormhole attacks and Sybil attacks an intruder can give misleading routes or even completely disrupt the network.
In addition to the complete overview of SB and their protocols given in \cite{granzer2010future}, various attacks are presented at all layers, including on wireless SACs such as network attacks (interception, modification, fabrication, and interruption) or device attacks (software, physical, and side-channel attacks).
Moreover, in addition to aforementioned attacks to sensor networks, papers \cite{sun2014review} and \cite{misra2010detection} present node replication and resource depletion attacks.
Embedded networks attacks are presented in \cite{praus2008secure} with attacks targeting SB devices such as code injection or reading their memory along with defence mechanisms for these attacks.
Several attacks and counter-measures applying to Advanced Metering Infrastructure (AMI) are presented in \cite{cleveland2008cyber}.
The main focus in this last publication is the smart grids to which SB can be directly connected, therefore many issues overlap on both context.
As an example, the author discuss the customer gateways which interconnect buildings to the grid and the various issues related to a physical or cyber compromise of these devices causing privacy, confidentiality, integrity, availability, and non-repudiation concerns. 

\subsection{Automation \& Management layer}

Attacks at the automation and management - or "backbone" - layer show many similarities with well known attacks on classic IT networks.

\subsubsection{Denial of Service Attacks}

Several publications are fully dedicated to DoS attacks at the automation and management layers \cite{granzer2008denial,judmayer2014security,kaur2015securing,antonini2014practical,vceleda2012flow,hager2012secure,mundt2016security,caviglione2015analysis,hill2004middleware,cleveland2008cyber}.
Two main type of DoS attacks can be conducted at this layer \cite{granzer2008denial}: host-based and network-based.
In host-based DoS, attackers target devices that are critical for the operation of the system (i.e., firewalls), devices that concentrate automation functionality (i.e., controllers), or interconnection devices (i.e., gateways, routers).
In network-based DoS, the target is the network itself, where the attackers try to waste the network bandwidth up to completely interrupting the communication between the devices located in the affected segment.
In terms of particular techniques to effectively launch a DoS, a prominent one was flooding \cite{vceleda2012flow}, \cite{kaur2015securing}, \cite{antonini2014practical}, \cite{judmayer2014security}, \cite{caviglione2015analysis}.
In this type of DoS attacks, an attacker uses weaknesses of the network or service where they become so overwhelmed with requests that they can no longer process genuine connection requests.
Examples include broadcasting "Who-Is" requests without specifying a device instance range limits in BACnet, or router advertisement flooding.

\subsubsection{Protocol-specific Attacks}

Many of the reviewed papers, as stated in the general analysis, discuss vulnerabilities and attacks that can be made against protocols like BACnet, KNX, and LonWorks \cite{granzer2006security}, \cite{granzer2010future}, \cite{granzer2010security}, \cite{vceleda2012flow}, \cite{judmayer2014security}.
Examples of attacks include \cite{vceleda2012flow}, which demonstrate a detection of Telnet brute force attacks and targeted attacks on building automation system networks, such as BACnet spoofing and Write-Property attack.
The former is similar to ARP spoofing, where compromised device generates fake I-Am-Router-To-Network messages of BACnet and forces other devices to send their messages via the device of an attacker.
Whereas, the paper \cite{judmayer2014security} discusses attacks made to KNX backbone network, such as tampering, ARP cache poisoning and replay attack.

In \cite{wendzel2012covert}, a group of researchers successfully achieve covert storage and covert timing channel attacks on a BACNet network.
This attack may evade most intrusion detection systems (passive wardens) and normalisers/firewalling systems (active wardens) as they do not usually take SB network protocols into account.
Covert channels are hidden communication channels which are not intended for information transfer at all.
The intention of a covert channel is to hide the existence of an information flow that possibly violates a system’s security policy.
In this publication, they describe three covert channel based on BACNet:
1) A message-type based covert storage channel, which basically consist in attributing a bit value to a certain type of message; 
2) Parameter based covert storage channel, which consist in using an unused flag in the headers of the communication to hide data; and 
3) A covert timing channel, which consist in making sense of the time in between messages on the network to recreate a hidden message.
A mitigation solution firewall based is then presented and will be detailed in the Defences section.

A good overview of the issues enabling attacks and vulnerabilities is provided in \cite{antonini2014security}.
Not surprisingly, one of the core issues is that these system were explicitly designed assuming the existence of physical barrier and isolation.
The constraints in computational resources of devices and the slow bandwidth of some protocols such as KNX (9.6kbps maximum) or LonWorks (1.25 Mbps) are also enabling factors facilitating different attacks.
Transmission of data in clear text in both wired and wireless network is common in building automation, without any integrity checking, and no security guarantees (except for network membership check).
Default configurations are also problematic.
BACNet as an example provides encryption capabilities but is configured to send messages in clear by default.
Modbus communications are not protected by any cryptographic primitive.
The researchers sum up that the most widespread solutions for building automation protocols do not include any explicit security feature, and in the vast majority of the cases, it is not possible to exploit the ones of the underlying transport layer.
Spoofing techniques are described as possible and easy, which allows to hijack and manipulate the traffic.
Simple sniffing techniques are also possible and can lead to a full compromise as for example KNX nodes authenticate commands with a clear-text password sent over the network.
The researchers then propose different counter measures that we will discuss in the Defences section.
Finally, \cite{praus2014identifying} conducted a world-wide scan looking for BACNet and KNX devices connected directly to the Internet, finding thousands of devices following these protocols potentially vulnerable. 

\subsubsection{Configuration and Legacy Vulnerabilities}

Some works reported attacks particularly exploiting how systems had been or can be configured in practice, or the fact that some systems may be outdated and implement older versions of protocols.
This includes the work reported in \cite{mundt2016security}, in which the authors investigated two SB where they found missing authentication, encryption and authorization mechanisms at the automation layer.
Besides, they reported no presence of firewalls or similar appliances.
It was also reported in \cite{hager2012secure} other potential vulnerabilities, including the disclosure of administrator passwords, account information, hacking of accounts, passwords or ciphers, etc. 
Insecure wireless, modems, remote support, and serial links as well as firewall misconfiguration in the context of both ICS and SB is discussed in \cite{boyes2015best}.
Another specific example of misconfiguration was illustrated by the authors of \cite{zito2014decoding}, when the authors discuss the case of an attacker who penetrated into a SB via default passwords that had not been changed by the control system installer. 
Finally, there may also be vulnerabilities coming from organisational aspects.
For instance, the authors of \cite{mundt2016security} found that there was no one really responsible for the security of the whole SB in the companies they investigated, no central user management, and no comprehensive overview of all installed systems.

\subsubsection{Privacy Attacks}

Researchers have also demonstrated several attacks that compromise the privacy of the building users (inhabitants or employees) \cite{wang2014non,caviglione2015analysis,mundt2016security}.
For instance, the authors of \cite{caviglione2015analysis} analyse surveillance attacks, where the attacker either directly requests sensor values from internet-connected smart buildings or exploits side channels such as timing information, electromagnetic or power consumption leaks, even sound may provide additional source of information to monitor the behaviour of building inhabitants or employees.
Another example is the research presented in \cite{wang2014non}, which discusses how time-series occupancy data collected for environmental control in smart buildings leaks privacy-sensitive information about the building users. Besides, the paper \cite{kumagai2004sensors}  discusses privacy concerns of new technological advances like GPS, RFID. 

\subsubsection{Others} 
Some other papers have considered the interplay between SB and other smart infrastructures like smart grids. For instance, \cite{wei2015security} discussed a guideline price manipulation attack, where malicious customers alter the guideline price (which is used to predict the future electricity price to guide electricity demand of customers) in the electricity market and delude scheduling decision of other customers. By this type of attack, a malicious customer can reduce his electricity cost while increasing electricity bill of other normal customers. In \cite{bo2014usable}, the authors surveyed and measured user's level of concern regarding the security in SBs. Finally, a number of papers highlight and discuss several potential attacks, vulnerabilities and security risks of SB from a more high level point of view, including \cite{fisk2012cyber,mylrea2015cyber,tankard2015secure,mansfield2015dangers,lee2013reliable,hutt1990control}.

\section{Defences}
We now focus and give further details on the defences discussed in the reviewed papers. Table \ref{table:defences} shows an overview of the papers categorized according to the SB layer and grouped according to their similarity in terms of the type of defence. We can observe that, 21/62 papers focus on field layer defences and 41/62 at automation and management layer.
At field layer, we find some defences against DoS (4/21), IDS (5/21), cryptography-based defences (6/21), and other approaches to security in the field layer (6/21).
Most of these defences are aimed at wireless communication at field layer, specifically for wireless sensor networks.
At automation and management layer, Intrusion detection and prevention defences are 12/41 papers, protocol-specific defences have 9/41 papers, network and data in transit defences 8/41 papers, cryptography-based defences 4/41, hardware/software-based defences 4/41, and 4/41 papers present guidelines and best practices. This clearly shows that the attention has focused on detecting and preventing particular types of attacks (including DoS) to make SB more resilient but without the need to actually touch the SB technologies themselves, but correcting security issues of SB protocols, as shown by protocol-specific defences having received a lot of attention to correct the weak security of the present solutions.
For instance, in 2010, BACNet was updated in the Addendum G, with the objective: \emph{``to provide peer entity, data origin, and operator authentication, as well as data confidentiality and integrity''} \cite{bacnetaddendumg}.

\begin{table}[!h]
\def\arraystretch{1.5}
\footnotesize
\begin{tabularx}{\textwidth}{|c|X|} 
\hline
\multicolumn{2}{| c |}{\begin{large}\textbf{Defences}\end{large}} \\
\hline
\multicolumn{2}{| c |}{\textbf{Field layer}} \\
\hline
Defences against DoS (4) &
\cite{sun2014review}
\cite{paridari2016cyber}
\cite{shen2009proactive}
\cite{muraleedharan2006cross}\\
\hline

Intrusion Detection (5) &
\cite{paridari2016cyber}
\cite{al2015anomaly}
\cite{dehestani2013robust}
\cite{luo2003multiagent}
\cite{li2013sketch}\\
\hline

Cryptography-based (6) & 
\cite{sparrow2015study}
\cite{garcia2009angel}
\cite{matthias2015study}
\cite{porambage2015group}
\cite{park2014lightweight}
\cite{seshabhattar2011hummingbird}\\
\hline

Others (6) &
\cite{misra2010detection}
\cite{durresi2009secure}
\cite{vukasinovic2012survey}
\cite{burke2014secure}
\cite{shang2014securing}
\cite{xiong2015wifi}\\
\hline

\multicolumn{2}{| c |}{\textbf{Automation \& Management Layer}} \\
\hline

Intrusion Detection \& Prevention (12) &
\cite{wendzel2012covert} 
\cite{wendzel2012covert1}
\cite{szlosarczyk2014towards}
\cite{vceleda2012flow}
\cite{meyn2009anomaly}
\cite{jaikumar2011detection}\cite{nanping2009application}
\cite{nanping2009optimal}
\cite{boyer2007improving}
\cite{wendzel2011secure}
\cite{yoon2011distributed}
\cite{kastner2011accessing} \\
\hline

Protocol-specific (9) &
\cite{judmayer2014security}
\cite{granzer2009securing}
\cite{antonini2013security}
\cite{pan2014anomaly}
\cite{tonejc2015visualizing}
\cite{holmberg2005ashrae}
\cite{kastner2004closer}
\cite{antonini2014security}
\cite{granzer2006security}\\
\hline

Network \& Data in transit (8) &
\cite{ungurean2016monitoring}
\cite{bordencea2011agent}
\cite{tong2013information}
\cite{lesueur2014palpable}
\cite{boussard2015software}
\cite{reinisch2008secure}
\cite{granzer2008denial}
\cite{granzer2010communication}\\
\hline

Cryptography-based (4) &
\cite{granzer2008key}
\cite{hager2012secure}
\cite{hong2012building}
\cite{wang2012wireless}\\
\hline

Hardware/Software-based (4) &
\cite{wang2015secure}
\cite{hernandez2015safir}
\cite{praus2008secure}
\cite{hill2004middleware}\\
\hline

Guidelines \& Best practices (4) &
\cite{zito2014decoding}
\cite{schneider2014bestpractices}
\cite{boyes2015best}
\cite{sutherland2015applying}\\
\hline
\end{tabularx}
\caption{Defences categorized according to the SB layer and grouped according to their similarity. \label{table:defences}}
\end{table}


\subsection{Field layer}
Defences at field layer focus mostly on methods to defend against wireless attacks. There were very few defences that talked about physical attacks. In terms of wireless attacks, the type of defences follow mainly the technologies suggested by studies like \cite{fantacci2013enabling}, mainly including specific defences against DoS, cryptography-based defences and intrusion detection. 
%
%

\subsubsection{Defences against DoS}

In \cite{sun2014review} a new architecture is presented allowing detection of compromised nodes in the network, resilience to sink hole attacks, a group based security scheme, and a secure key establishment for WSN.
The researchers present limitations found in all WSN such as the node and network limitations to deploy security including limited energy, storage room, poor computer ability, highly dynamic network topology, and low bandwidth.
The different attacks and threats are also presented such as physical attacks, node clones, and numerous DoS attacks. 
At the physical level, counter measures include access prohibition to avoid physical attacks, which is often infeasible, and encryption.
At the data link layer, where attacks are generally traffic manipulation and spoofing, the defence moves towards detection, with identity and behavioural detection that can trigger mechanisms to exclude attacking nodes from the system.
At the network layer, the attacks can absorb the traffic, use Man in the Middle techniques and control the traffic flow.
To counter these attacks, routing access restriction, false routing information detection, and wormhole detection can be set up.
At the last layer, the application layer, the possible attacks target mainly the application semantics.
To prevent this, defences protecting the data integrity and confidentiality can be set up, possibly using cryptography and hashing mechanisms. 
%
To prevent DoS aiming the WSN network \cite{muraleedharan2006cross} make use of Swarm intelligence, derived from insects intelligence, to detect the possible routing and the best routing performances.
An initial set of agents traverse the network through nodes in a random manner, and once they reach their destination they deposit the digital equivalent of pheromone trail as a means of communicating indirectly with the other agents.
The pheromone accumulation is proportional to the number of agents travelling between two nodes during a complete iteration.
This solution increases performance, redundant routing path, and therefore provides denial of service resilience. 
Finally, in 
WSN, compromised nodes are one of the main threats.
Once a rogue node is in the network, defences that do not consider this eventuality might fail to provide any security at all.
For this reason, in \cite{shen2009proactive}, a Proactive Trust Management System (PTMS) has been designed to give tags of trustworthiness to sensors, specially tailored for resource constrained devices, therefore suitable for the IoT and WSN generally.
With the Quality of Information (QoI) tag, it maintains and aggregates a notion of reputation of every node in the WSN that can be used to trigger security measures.


\subsubsection{Intrusion detection}
An example is presented in \cite{paridari2016cyber}, where the researchers attack temperature sensors to provide false data to the energy management system, leading to an energy degradation costing thousands per year.
In this paper, the researcher adopt an interesting physical approach to detect anomalies and outliers.
To summarise it, they use the physical measurements to detect anomalies in the system when the measurement do not make sense. In a similar way, but focusing on faults instead of necessarily malicious intrusions, \cite{dehestani2013robust} developed a fault detection system using Support Vector Machines (SVM) and Artificial Neural Networks (ANN). 
%
%
In \cite{al2015anomaly}, new DoS risks and threats are presented, including in addition jamming, delay attacks, pulse DoS, and NWK knockdown.
Their approach is an intrusion detection system using anomaly behaviour detection, allowing them to detect both known and unknown attacks.
Capturing packets from the network to feed a database that is used to learn what is a normal activity to then detect abnormal activity with good results (95\%+ in all their scenarios).
%
Another paper using anomaly detection is presented in \cite{li2013sketch}, also with good results in terms of accuracy along with low false alarm ratio. Finally, \cite{luo2003multiagent} presents development of a multiagent multisensor-based security system for intelligent buildings.
They integrated fire detection/diagnosis agent, intruder detection/diagnosis agent, environment detection/diagnosis agent (humidity, gas, temperature detection) and power detection/diagnosis agent.
These subsystems are all ruled by four methods, namely: adaptive fusion method, rule-based method, redundancy management method and static method.

\subsubsection{Cryptography-based defences}
Classic IT security measures are being tested in the specific context of WSN, such as the use of software authenticated encryption with associated data (AEAD) with cipher block chaining and message authentication code (CCM) in \cite{sparrow2015study}.
The results suggest that the current standardised AEAD algorithms are not as suited in comparison to adjustable AEAD if the packet throughput is a priority. %
%
Another security mechanism used in IT networks being adapted at field level in WSN is the use of key infrastructure.
In \cite{garcia2009angel}, researchers present their security architecture that ensures the secure deployment and operation of ANGEL systems, ZigBee based sensors targeted at healthcare and assisted living.
%
In \cite{seshabhattar2011hummingbird} a mutually authenticated key establishment procedure named HummingBird Key Establishment (HBKE) to secure key agreement between an initiator and responder device is presented.
Once again, this solution provides highly efficient mutual authentication and privacy with little communication overhead using AES.
This solution seems to provide high security with smaller hardware sizes, faster operation, and smaller power consumption.
%
In \cite{porambage2015group}, researchers present a group key establishment in WSN for IoT applications that provide resilience to DoS attacks, with a network comprising compromised hosts, replay attacks, and MitM techniques.
The main objective of this solution is to secure multicast communications.
This technique makes use of elliptic curve cryptography (ECC) and has characteristics suitable for IoT devices.
%
A group of researchers attempted in \cite{park2014lightweight} to delegate the DTLS handshake at the beginning of an encrypted transmission to increase security.
This method uses a Secure Service Manager (SSM) protected against spoofing, provides semi end to end security, DoS resilience as DTLS handshakes can be resource intensive and are in this method delegated, and remove single points of failures.
In
 \cite{matthias2015study}, the researchers investigate the impact of including low cost security solutions into the communication scheme on latency, power, and memory management.
The security solution they benchmark makes use of AES-CCM which is one of the favourites encryption mechanisms in WSN accordingly to our study, along with RC4 in some cases for its performances and low computational cost. The study shows a small increase of memory used of 4 to 6 percent with a higher impact on battery power as a battery powered device would be able to run only for 70 days in their tests.
Finally, \cite{xiong2015wifi} studies privacy protection in a smart building from Wi-Fi point of view. The authors propose a solution to prevent the leakage of Wi-Fi signal based on FSS (Frequency Selective Surface) with dual-band stop behaviour. 

\subsubsection{Others}
%
%
In \cite{misra2010detection}, the authors solve the problem of identifying hosts not with IP or MAC addresses as these information are not secure and can be spoofed, but instead by creating wireless signal fingerprint - or "signalprint" - and use this unique signalprint to identify a host.
In the system every wireless sensor participates to the task of identification.
Each of the wireless sensors create a signalprint of the other sensors at reach and each nodes compare the signal to the recorded signalprint to certify that the sensor communicating is legitimate.
This approach is hard to spoof, has a strong correlation with physical location, and small variation over moderate amount of time.
The more sensors participate in this system, the most efficient this system becomes.
Limitations include imperfection when the battery level varies which change the signal, moving sensors are not suitable for this solution, and the sensors have to communicate to identify another sensor, leading to increased network traffic and therefore battery usage.


Another interesting and uncommon approach is presented in \cite{durresi2009secure} where the researcher attempt to make use of cell phones that they include in the WSN to obtain relevant information in real time, increase communication capacities as cell phones could communicate both with the sensor network and the cellular network, which can also be used to increase authentication and security.
In this solution, the authors develop gateways to be bridges between the cell phones and the sensor network.
They use symmetric and asymmetric encryption mechanisms to secure the communications. They are also capable of detecting and revoking captured gateways. Another paper also explored the possibility to use mobile agents such as cell phones in WSN to improve them on different aspects such as autonomicity, security, adaptability, and recovery \cite{vukasinovic2012survey}.


Finally there have been attempts to solve a number of problems in SB sensor networks by making use of Named Data Networking (NDN) \cite{shang2014securing,burke2014secure}.
By using a hierarchical namespace for data, encryption keys, access control lists, and encryption-based access controls, the researchers have built a testbed that collects sensing data from industry-standard components, and implemented a prototype of access controlled by their system.
Instead of communicating in a classic way, all the data is centralised, categorised, and made available in the NDN network using their own security mechanisms, which eliminates numerous problems. 


\subsection{Automation \& Management Layers}

We find in the defences for the management and automation layer numerous techniques adapted from classic IT networks for the specific case of SB.
These techniques include DoS detection, prevention, and recovery, firewalls, encryption along with secure key establishment, access control management, single point of failure prevention, intrusion detection systems and anomaly detection, monitoring and logging, and various guidelines and best practices.

%



\subsubsection{Intrusion detection and prevention}

The authors of \cite{wendzel2012covert} developed detection and mitigation tools for covert storage and covert timing channel attacks with a BACNet Firewall Router (BFR) - which is available for download. 
Other papers propose traffic normalization methods like \cite{wendzel2012covert1} against covert channels and side channels, and \cite{szlosarczyk2014towards} exemplified using BACnet. 
%
%
Other works focused more on detection. In \cite{vceleda2012flow}, the authors based on data flows to parse, and store BACNet headers and messages to then detect flooding DoS attacks and other known BACNet attacks such as spoofing, write property attack, and disabling network connection. In \cite{yoon2011distributed}, the authors include DDoS detection as part of their security infrastructure for SB. The authors of \cite{meyn2009anomaly} demonstrate techniques to deal with detection of anomalies in interconnected and dynamic systems, where the authors successfully adopted projected Markov models.
In \cite{jaikumar2011detection}, the authors show a way of learning method for detection of anomalies using multi-modal smart environment sensor data in patterns of human behaviour. Finally, the work presented in \cite{nanping2009application,nanping2009optimal} makes use of an IDS that redirects the attacking traffic to a honeypot using Honeynet which allows to study real attacks against SB. 
Finally, some other works proposed different types of gateways for SB. In \cite{boyer2007improving}, a gateway relying mainly on XML exchanges over HTTPS is proposed to improve authentication. In this project a set of redundant authentication gateways are developed that manage the access of users to end points, which can be used to enforce a security policy. 
%
The work presented in \cite{wendzel2011secure} present some similarities in its concept with \cite{boyer2007improving}.
The authors have developed a multilayer architecture for building automation that allows remote management and enables using components from different manufacturers with backward compatibility.
Their solution presents an interface using secure event-handling and role based access control. This system also acts as a gateway in between the users and the end hardware, checking their authorisation and granting or denying access accordingly to the rules defined.
The researchers point out that this solution does not solve the security problems of those components, but focuses on the security of higher level interfaces and on the unification of security features. The authors of \cite{kastner2011accessing} present an inter-protocol gateway prototype using BACnet over Web Service for KNX networks, allowing these two technology to interact together and with other applications in classic IT network.




\subsubsection{Network \& data in transit defences}

In \cite{bordencea2011agent}, the researchers developed a software agent based system providing adaptation and fault tolerance allowing a system to continue to function in presence of access point failure or defective sensors.
Their solution allocate sensors to specific access points, providing virtual redundancy, therefore increasing reliability. 
%
An analysis of the information flow in SB is presented in \cite{tong2013information} along with a security model to secure the information during its transmission.
Starting by sorting the data by type, level and category (confidential or non-confidential), they create a package's security label that is then used by their system to secure the transmission.
Their system is tested in two scenarios of attacks: malicious control of a node and confidential information leak.
The two attacks are mitigated by their system thank to the detection of mismatch in the security data label. 
%
%
The paper \cite{lesueur2014palpable} proposes a solution for the dissemination of private data by introducing novel access control model called Tuple-Based Access Control (TBAC) that tracks sensor information flows. 
%
The paper \cite{boussard2015software} presents a software-based solution, built on Software Defined Networking (SDN) principles, for interconnection of devices of smart environments depending on requests and expectations of end users. 
%
A project supporting various protocols and incorporating an abstracted security layer has been developed \cite{reinisch2008secure}.
A proof of concept device supporting both KNX and BACNet has been created for this project. A monitoring and control system for SB was developed based on the OPC UA specification in \cite{ungurean2016monitoring}, integrating ModBus and BACNet fieldbuses and with connected smart plugs using a ZigBee to ModBus gateway, allowing the local and remote control and monitoring of the devices. 
In \cite{granzer2010communication}, they propose of set requirements for secure communication in SB, including: entity authentication, authorisation, secured channel, data integrity, data origin authentication, data freshness, data confidentiality, data availability; and secondarily, anonymity, and auditability.
In this same paper, the researchers develop a solution that allows abstracting the different protocols used in SB to implement security using gateways in charge of the security operations. 
Finally, a group of researchers attempt in \cite{granzer2008denial} to solve the difference of computational power between automation system devices and potential attackers which could lead to a resource exhaustion DoS by developing a system where the client has to solve a complex puzzle whose solution can be simply compared to an expected result without calculation in the device.

\subsubsection{Protocol-specific defences}
Some protocol-specific defences have been developed besides adaptations of IT security measures.
Starting with KNX, 
  researchers compared two security extensions for KNX/IP (KNX/IP Secure and the generic security concept) in terms of security \cite{judmayer2014security}, concluding that further evaluation and testing is required before deploying widely any security extension, and that in the meantime, 
 classic IT security measures such as strict physical and logical network separation, and usage of VPN should be used.
%
In \cite{granzer2009securing}, the lack of security in KNX is improved with the use of an online key server and elliptic curve cryptography (ECC).
%
In \cite{antonini2013security}, the authors propose a multiparty key agreement scheme equivalent to Diffie-Hellman problem, however, their solution seems resource demanding as it requires one minute of computation on a KNX network. 
Numerous security challenges in SB are discussed in \cite{antonini2014security}, along with possible counter measures such as packet filtering or proxies-filters to strengthen KNX security.
\cite{granzer2006security} proposes a security extension of EIB/KNX (named EIBSec) enforcing data integrity, confidentiality and freshness, as well as authentication to provide secure process data and management communications.


Regarding BACNet, in \cite{pan2014anomaly} an IDS is developed for the protocol BACNet which allows its deployment without modification of the devices, allowing rapid deployment on running systems.
Their system show good performances at detection of known attacks such as who-is/who-has attacks, write property attacks, initialize routing table attacks, reinitialise device attacks, device communication control attacks, and flooding conformed service attacks.
%
To help the manual detection of anomalies or attacks, researchers have developed a project to visualise and analyse data from BACNet \cite{tonejc2015visualizing}.
This tool is aimed at system operators, it provide graphs, network message flow maps and more features to help in the real time security assessment.
Using it, people without experience in BAS could identify anomalies at 72.5\% and 52.5\% correctness.
%
A set of twelve different security goals to achieve for BACNet are discussed in \cite{holmberg2005ashrae}, however we would like to point out that this paper is dated from 2005 and BACNet has known several improvement since its publication, an entirely new standard is today available.

Finally, \cite{kastner2004closer} discusses popular protocols in SB (BACnet, LonWorks, EIB) and compares technical aspects of building and home automations, like platforms, network scalability, interoperability, security, reliability of transport, and general conditions for development, like tools, costs, licence policies.

\subsubsection{Cryptography-based defences}
A key set management system using multiple key servers to avoid single points of failure specific for SB has been developed in \cite{granzer2008key}. 
%
Aiming at improving the security and quality of service in these systems, the authors of \cite{hager2012secure} implement encryption for data at rest and in movement, authentication of participants, key exchange management and role based access control.
%
In order to provide remote control and access, SB are increasingly embedding web servers in their devices and software.
Consequently, the security of web interfaces of SB is also of prime importance.
In \cite{hong2012building}, a web service allowing to control lighting in a building is developed.
The researchers use corrected block TEA encryption, hashing functions, authentication, and logging to provide security to their platform.
In \cite{wang2012wireless}, several cryptographic tools such as RAS are used to ensure data transmission security and reliability. 

\subsubsection{Hardware/software-based defences}
\cite{wang2015secure}\cite{hernandez2015safir}\cite{praus2008secure}
An ARM-compliant framework has been created specifically for IoT-enabled devices that can be found in SB \cite{hernandez2015safir} with strong emphasis on security and privacy aspects to be used in smart buildings. 
%
Another research project working is presented in \cite{praus2008secure} to allow the execution of untrusted and possible intentional malicious code to be executed securely on a low end embedded system.
They present a proof of concept and evaluation on a SB.
To achieve a highly secure and trustworthy control system that can be used without major disruption to the existing infrastructure the authors of \cite{wang2015secure} designed a security/safety modeling framework for building controls.
It comprises an RTOS (Real-Time Operating System), which may ensure that the attacks made towards embedded controllers (which are widely used in BAS) will be stopped. In \cite{hill2004middleware}, a system of middleware boxes is presented to mainly manage access control but also provides through a system of redundancy and load balancing some DoS resilience.

%
%

\subsubsection{Guidelines \& best practices}

Besides the technical discussions and other experiments, guidelines and best practices are available in the literature. In \cite{zito2014decoding}, the potential disastrous consequences of a SB failing are evoked along with potential counter measures and best practices.
%
These advices include keeping the system up to date, adding monitoring and logging features, planning backups, having clear and least privilege access for users, and redundancy where possible, depending on the criticality of the system.
An interesting idea is also given in this paper, to make the manufacturers responsible to send the vulnerability notification to their customer. 
%
Schneider electric also published five best practices guide to improve SB cybersecurity \cite{schneider2014bestpractices}.
The guidelines are to: 1) change default passwords; 2) change default ports of services including web interfaces; 3) disable unused interfaces such as USB; 4) add network security such as a firewall closing unnecessary ports, fragment the network, enforce least privilege access for users, remove accounts when needed, and change account when an employee's role change; and 5) apply the patches and updates as soon as possible.
%
The authors of \cite{sutherland2015applying} explore the application of traditional digital forensics practices by applying established good practices guidelines to the field of building automation.
They examine the application of the UK Association of Chief Police Officers (ACPO) guidelines for digital evidence, and attempt to identify the challenges and gaps that arise in the process, procedures, and available tools.
Finally, in
\cite{boyes2015best}, the authors 
present software trustworthiness, and how to measure the level of trustworthiness required for different elements accordingly to their importance in the system, from paramount to ancillary.
A vulnerability assessment methodology is provided with the different relationships between unities involved in a system.


\section{Open Challenges}
Based on the results of our systematic literature review, we identified a set of open challenges that we believe deserve further attention from the research community on SB security. Next, we discuss them in detail. 

\subsection{The Growing Complexity of Smart Buildings}

As more sensors and actuators are being included in Smart Buildings, their complexity is growing over time.
A clear example of this growing complexity is the number of different protocols that have been introduced in recent years, and that research papers have started to consider, both wired and wireless, as we have shown in this systematic literature review (cf. Section \ref{sec:analysis} and in particular Figures \ref{fig:protocols} and \ref{fig:protocolsbar}).
This also means that research efforts on the security of smart buildings, even if they have clearly increased over the last 2-3 years face a growing attack surface. Also, more complexity and protocols means that more undesired interactions between different components or protocols could lead to further security vulnerabilities, and systematic ways of assessing and resolving those potential security vulnerabilities are a very interesting area for future research. This is even more critical given the push for SB as a strategic point in the Smart Cities paradigm \cite{gasco2018building}, which requires further integration, hence increasing complexity more, between SB infrastructures and other city resources \cite{sita2014knx}. 

A very important part of this challenge is how to ensure secure interoperability between different SB protocols. 
%
%
As an example, KNX devices and network messages cannot be understood by a device using BACNet.
Some systems of gateways are available to make different system relying on different protocols to collaborate.
A good example of an inter-protocol gateway prototype is presented in \cite{kastner2011accessing} using BACnet over Web Service for KNX networks, allowing these two technology to interact together and with other applications in classic IT network.
Other gateways to bridge protocols exists, such as BACNet to and from ModBus, and many others are theoretically possible.
The difficulty to have fully compatible implementations of the one same protocol, guaranteeing that two products from different manufacturers can interact with each other is already challenging.
This difficulty made the main protocol authorities to have product certification programs, such as the BACNet Testing Laboratory (\url{http://www.bacnetinternational.net/btl/}), the KNX certified products (\url{https://www.knx.org/ae/knx/knx-products/knx-certified-devices/}), LonMark certified products for LonWorks (\url{http://www.lonmark.org/product/}).
Same certification exist for wireless protocols used mainly at field level in WSN such as ZigBee as an example.

\subsection{Lack of Empirical Evaluations}

Although we found a number of articles that contained evaluations, and many of them seemed to increasingly consider testbeds and real systems for the evaluations, there was a relatively high number of articles (even in recent years) that did not include proper and systematic evaluation --- cf. Section \ref{sec:analysis} and in particular Figures \ref{fig:Evaluated} and \ref{fig:evaluatedbar}.
This could be due to the inherent difficulties involved in having realistic and meaningful evaluations.
SB testbeds are normally expensive and not every research group or university can afford one, and real systems often require very good partnerships with industry, and even in that case it is difficult to convince organisations to lend their infrastructures for testing. Even when they do, because of the type of system, it may be that not all types of evaluations and security testing can be realistically conducted. For instance, well-known assurance techniques (cf. \cite{such} for a review of assurance techniques) such as penetration testing, formal verification, fuzzing and others  may have different requirements and potential impacts in an infrastructure. As any other CPS like Industrial Control Systems \cite{knowles2015assurance}, SB also have the problem that testing might need to be conducted in the operational infrastructure with potential devastating impact should anything go wrong. 
We however observed an increasing number of testbeds in universities, and the ability to share them across universities seem crucial here to further progress on this matter. 

\subsection{Little consideration of social issues and human factors}
We did not find a single paper that considered attacks, vulnerabilities or defences from a more \emph{social} angle.
Only one paper that discussed the lack of secure configurations, the various misuses, and persons responsible for them \cite{caviglione2015analysis} concluded that security problems emerge mainly from people-related issues, including: 1) vendors not integrating security by lack of know-how, 2) customers lacking security awareness, and 3) operators being non-technical people, therefore also lacking know-how; all this being catalysed by a poor optimisation during design and project deployment.
In contrast, issues like social engineering and phishing in SB have largely escaped any analysis or further consideration, let alone solutions and defences against these threats. Only one survey goes in this direction \cite{bo2014usable}, with 
the results suggesting that uses do care about security in SB but that current security mechanisms are not satisfying. In particular, several authentication methods were surveyed with participants preferring the pattern locker. A very interesting future research line is therefore creating usable security mechanisms for SB. This could draw from the extensive usable security literature to devise (cf. \cite{cranor2005security} for instance) the best methods to design, implement and test security mechanisms for SB that are easy to use by users yet provide a high level of SB security.

\subsection{Responsibility, liability, and lifecycle issues}

SB are various and require different sets of skills to be operated correctly.
For this reason, SB are the final product of numerous different contractors working on their respective subsystems most of the time connected to a central controlling unit, either a controller or a supervision server. We only found one paper that briefly touched upon this important problem \cite{mundt2016security}. In particular, the authors found that there was no one really responsible
for the security for the whole SB they were considering in their assessment. In particular, security was a matter of the IT department but there was no communication between
facility managers and the IT department. Facility managers in turn were not responsible for the SB security. In fact, security was neither a requirement when the SB was commissioned nor in the invitation to bid, there was not a comprehensive overview of all installed systems, different
companies installed different SB devices at different times, active components were replaced
during maintenance without updating the documentations, and there was no central user management for the whole SB. This can clearly cause problems in terms of liability and responsibility, beyond obviously making the security of the whole SB very poor. Therefore, a very attractive line of future research is to study and better integrate organizational and supply-chain factors towards more secure SB along their lifecycle. 

\subsection{Lack of consideration of composite vulnerabilities}
Very few papers from those that we reviewed considered the issue of the composition of vulnerabilities, which could actually facilitate attacks penetrating a SB across the 3 different layers of SB, even potentially making the jump and compromising the IT network the SB may be connected to. This type of vulnerabilities are known to occur in other CPS \cite{ciholascomposite}, but further work is needed to understand them in the context of SB. For instance, \cite{mundt2016security} already showed that once a SB was compromised physically at the field layer, they could also send messages to and compromise the automation layer. Future work should provide a better understanding of the interconnections and dependencies that single vulnerabilities can have. This will enable the development of methodologies to detect and chain single vulnerabilities within a SB, allowing to recognise the different patterns of interactions and/or dependencies between single vulnerabilities, which can then reveal composite vulnerabilities. The efforts in this process should be focusing on having a formal representation of the framework and a clear methodology to go from single vulnerabilities as input and produce a set of composite vulnerabilities, including and accounting for the severity of such composite vulnerabilities in terms of their potential impact on SB. The ultimate aim should be to explore to what extent composite vulnerability detection and analysis could be automated, and to what extent and through what kind of (potentially novel) defences this type of vulnerabilities could be mitigated.


\subsection{Making Smart Buildings \emph{Smarter}}
We observed that very few papers were taking advantage of AI techniques to design smart defences for SB. We believe there is a very good opportunity for SB security research based on AI techniques. Note these AI techniques would include not only data-driven techniques like Machine Learning but also knowledge-based techniques such as normative systems, which have been successfully used to develop intelligent security and privacy methods in other domains \cite{such2016intelligent,such2017privacy}. Examples of the use of ML techniques include intelligent intrusion detection and prevention methods but also automatic ways to configure permissions of the different components in SB. Similar research has already been shown useful in other domains like Mobile App permission management \cite{olejnik2017smarper} and Social Media sharing settings \cite{misra2017pacman}. Importantly, any ML-based methods need to be engineered considering adversarial cases \cite{goodfellow2018making}. Examples of the use of knowledge-based AI techniques include the use of norms, which have been extensively studied in recent years particularly as a way of limiting the autonomy of autonomous and intelligent systems to adhere to acceptable behaviors \cite{criado2011open}. Norms are usually defined formally using deontic logic to state obligations, prohibitions, and permissions, providing a rich framework to express context-dependent policies based on Contextual Integrity \cite{nissenbaum2004privacy}. 
Other AI techniques that could be used include negotiation \cite{baarslag2016negotiation,such2016privacy} for different SB components to negotiate access control, and computational trust \cite{pinyol2013computational} to select and only share data with trustworthy and privacy-respecting SB components and devices.

\subsection{Securing Internet-facing Smart Buildings}

Many SB are directly facing the internet which exposes these systems to remote attacks  \cite{praus2014identifying}.
Different elements can be Internet-facing, including the centralised supervision server, the SB controllers, human-SB interfaces, individual sensors or actuators, and other elements.
Several understandable reasons can lead to architect a SB with Internet-facing devices.
One of the main reason is to allow remote maintenance and troubleshooting, which allows to save time and money by doing operations over the Internet instead of sending an engineer on site.
In the case of geographically large or multiple building SB, each building, floor, or more generally separate area can be connected to a single device or server facing the Internet, which then allows a centralised solution to connect and manage them all.
In a recent papers of those we reviewed, researchers demonstrated this problem by using the Internet-wide network scanning platform Shodan (\url{http://www.shodanhq.com/}) to detect SBs using BACNet or KNX/IP directly connected to the Internet around the globe \cite{praus2014identifying}.
Having Internet-facing services exposes SB to a wide range of remote attacks having diverse goals.
SB devices and software often lack basic security mechanisms which could make them prime targets for attacks as shown in this systematic literature review. Some researchers even foresee the upcoming uprising of SB botnets \cite{caviglione2015analysis} \cite{sutherland2015applying} for instance. The challenge is therefore to allow remote access to SB without allowing potential attackers to easily detect or exploit them.
In addition, further research is required to assess and measure accurately the security of Internet-facing SB more thoroughly than what was done in \cite{praus2014identifying}, since simple detection of such systems does not inform us on their security posture and devices detected could even be a Honeypot.

\bibliographystyle{elsarticle-num-names}
\bibliography{The_Security_of_Smart_Buildings_-_A_systematic_literature_review} 

\end{document}